\documentclass[a4paper,11pt]{article}
\pdfoutput=1 

\usepackage{jheppub} 

\usepackage[T1]{fontenc} 
\usepackage{tikz,lipsum,lmodern}
\usepackage[most]{tcolorbox}

\title{\boldmath Emergent gravity from patterns in natural numbers}

\author{Atreya Chatterjee}

\affiliation{Harish-Chandra Research Institute, Chhatnag Road, Jhunsi, Prayagraj-211019, India}

\emailAdd{atreyachatterjee@hri.res.in}

\abstract{There is natural association of entropy with gravitational systems on one hand and partition of natural numbers on the other hand. We show that given a partition of natural numbers, it is possible to directly associate a metric with it. Gravity emerges from patterns in partition. In the process, metric and matter is unified into a fundamental notion of partition. More precisely, we find a common origin of Schwarzschild metric on one hand and black-hole entropy on the other hand. It immediately implies that information and metric are one and the same and any change in information is stored as change in metric. Thus gravitational radiation carries black-hole entropy worth of information. There are three novel experimental predictions. First, we can retrieve information from the gravitational radiation emitted during merger. Second, if radiation with right information is sent in, black hole absorbs information and decays instead of increasing in mass. This is reverse process of black hole formation. Third, till now only known way of observing nature is through radiation and fields measured far from the source. There is a completely new way of seeing nature if we can capture the whole partition of the source in one go.}

\begin{document} 
\maketitle
\flushbottom

\section{Motivation}

Emergent gravity means two things in this work. First it means equations
of gravity are emergent. It is more useful to call this emergent causality
as defined below. Second it means metric and matter emerge from some
fundamental degrees of freedom. In other words, unification of metric
and matter. There are other attempts at emergent gravity for example
using Bose-Einstein condensates \citep{Jannes:2009yr,Finazzi:2011zw},
entropic gravity \citep{Vancea:2010vf,Verlinde:2010hp,Verlinde:2016toy},
non-commutative geometry \citep{Yang:2007qx,Rivelles:2011gq}, quantum
computation \citep{Lloyd:2005js}, matrix models \citep{Steinacker:2010rh},
holographic models \citep{Almheiri:2014lwa,Harlow:2016vwg,Horowitz:2006ct,Kovtun:2004de}
and many other models \citep{Heckman:2011qu,Carlip:2012wa,Marolf:2014yga}.
This work is different from all other models as this is the first
example of emergent causality.

Causality ensures that events on one time-slice is completely determined
by events on another time-slice. At the heart of causality lies differential
equations which relate data on two slices. So by \emph{causality}
we mean \emph{a system described by differential equation}. This is
much weaker than lightcone causality. There are at least two handicaps with causality.
One it requires choice of initial and boundary condition from outside
\citep{Giddings:1988cx,Hartle:1983ai,Hartle:1986eu,Wudka:1987ef}.
Second, space of theory where differential equations live is \emph{separate}
and \emph{independent} from the phase space. Space of theory reads
coordinates of phase space as input, processes them and returns new
coordinates to the phase space. As a result, it requires storing,
processing and carrying forward information from one slice to another.
For example, be it a classical trajectory or some quantum process,
for all practical purposes we work with finite precision. But how
does nature process and keep track of large (infinite) number of decimal
places? Not only that but also the processes (eg. Feynman diagrams) are
infinitely numerous. In quantum gravity it is expected that information
is not just a label, but is one and the same as spacetime. Above problem
boils down to how information encoded in numbers (input and output
of theory) appear as physical space and time (phase space)? We see
spacetime and not numbers. \textit{Independence} means that phase space does
not restrict the space of theory. For example, there could have been
other consistent gravity, like higher curvature gravity describing
our spacetime. It is difficult to argue whether these handicaps are
problem or not but it appears to be inefficient.

Weaker definition of causality results in stronger definition of its
complement. By \textit{emergent causality} we mean a system which
is not governed by differential equation. Patterns emerge only over
some scale. One classic example is distribution of primes. Given $n^{th}$
prime, there is no formula which land us exactly on the next prime
but there are formulas which take us close to the next prime. So pattern
is emergent. $n^{th}$ prime has some but not all information about
other primes. However the system is deterministic as one can list
all the primes using the definition. Such a system is free from both
the handicap. There is no choice in the initial condition. For example,
the first prime is built into the definition of the primes. Second,
emergent pattern in density of primes is also invisibly built into
the definition of primes. Space of theory is integrated into the phase
space.

Equations of motions in physics relate field to its source. Einstein\textquoteright s
equations relate metric to energy-momentum tensor. However it is still
a step away from unification because energy-momentum tensor and metric
have independent definition. A system having emergent causality is
not governed by equations of motion. Thus such a system must relate
metric and stress-tensor in a more fundamental way. Presumably metric
and matter will be unified into one quantity which can be interpreted
either way depending on the way we see it. We will call this \textit{emergent
matter} and \textit{emergent spacetime}. There is another intuitive
way to see this. If fundamentally matter is nothing but information
associated with its blackhole entropy. Then spacetime and information
must be one and the same in quantum gravity implies that metric and
matter must be unified. Things will become clearer as we proceed.

\begin{figure}[tph]
\textcolor{black}{\centering \includegraphics[bb=20bp 5bp 604bp 403bp,clip,scale=0.65]{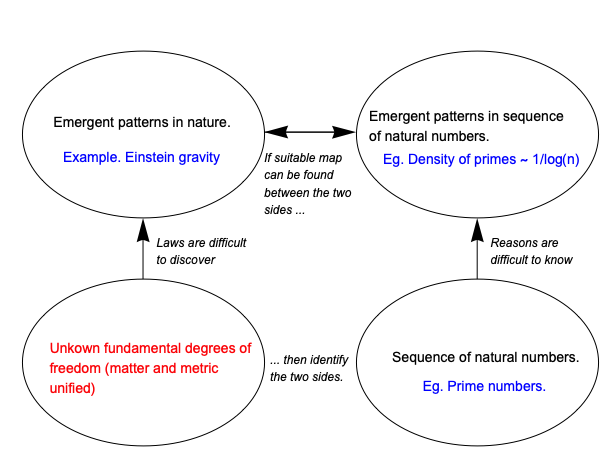}\caption{\label{fig:motivation}{\footnotesize{}Above illustration describes
the motivation. Goal is to directly map patterns in sequences of natural
numbers to patterns in nature like gravity. Then claim that underlying
degrees of freedom in nature is same as the sequence of natural numbers. }}
}
\end{figure}

In physics, main source of difficulty lies in handling interaction.
One main motivation behind this approach is that emergent patterns
have inherent interactions. For example, a prime has some some influence
over distribution of other primes. Somehow primes interact with each
other. Exact reasons are difficult to know. Natural numbers is the
simplest place where emergent patterns can be found. Underlying reasons
are often difficult to know. Similarly, we find emergent patterns
in nature, like gravity. Underlying fundamental laws are difficult
to discover. Goal is to directly map patterns in sequences of natural
numbers to patterns in nature. Then claim that underlying degrees
of freedom in nature is same as the sequence of natural numbers. Figure
\eqref{fig:motivation} illustrates the idea.

\textcolor{black}{}

\section{Introduction}

\begin{figure}[tph]
\textcolor{black}{\centering \includegraphics[bb=6bp 80bp 538bp 294bp,clip,scale=0.7]{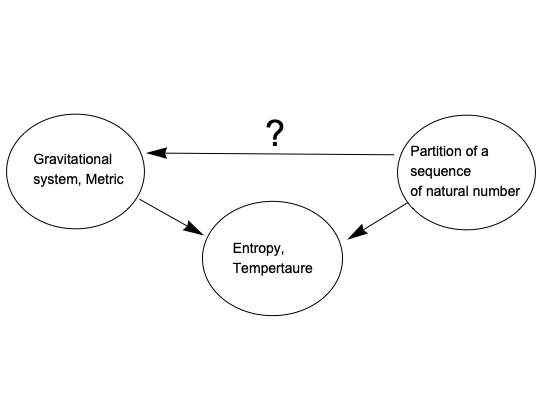}\caption{\label{fig:partition-metric}{\footnotesize{}Since there is natural
association of entropy and temperature with gravitational systems
on one hand and partition of natural numbers on the other hand. Above
illustration asks whether given a partition of sequence of natural
numbers, is it possible to associate a metric with it? }}
}
\end{figure}

\textcolor{black}{Given a configuration of black holes, there is a
natural definition of entropy. On the other hand, given a sequence
of natural numbers, using partition one can define entropy. Since
entropy is so naturally connected with the two fields, we explore
the following question in this paper. Given a sequence of natural
numbers and partition, is there a direct way to get an emergent metric?
Figure \eqref{fig:partition-metric} shows the idea.}

\textcolor{black}{}
Partition is the number of different ways of expressing a natural
number as sum of smaller natural numbers. Consider following partitions\textcolor{black}{
\begin{eqnarray*}
P\{2\} & = & \left\{ \{2\},\{1,1\}\right\} \\
P\{3\} & = & \left\{ \{3\},\{2,1\},\{1,1,1\}\right\} 
\end{eqnarray*}
}One can think of these as degeneracy. $\{2\},\{1,1\}$ will be called
\emph{parts} and their set \textcolor{black}{$P\{2\}=\left\{ \{2\},\{1,1\}\right\} $
will be called }\textcolor{black}{\emph{partition}}\textcolor{black}{.
We will use the terms partition and degeneracy interchangeably. Number
of parts will be denoted by $P(n)=|P\{n\}|$ .} Now imagine combining
$P\{2\},P\{3\}$ to form
\begin{eqnarray*}
P\{5\} & = & \left\{ \{5\},\{4,1\},\{3,2\},\{3,1,1\},\{2,2,1\},\{2,1,1,1\},\{1,1,1,1,1\}\right\} 
\end{eqnarray*}

One part each from $P\{2\}$ and $P\{3\}$ combine (denoted by $\bullet$)
to form a part of $P\{5\}$. For example, some of them are 
\begin{eqnarray}
\{2\}\bullet\{3\} & \rightarrow & \{5\}\label{eq:3pt}\\
\{2\}\bullet\{3\} & \rightarrow & \{3,2\}\label{eq:4pt}\\
\{2\}\bullet\{2,1\} & \rightarrow & \{4,1\}\label{eq:5pt}\\
\{2\}\bullet\{2,1\} & \rightarrow & \{3,2\}\nonumber \\
\{2\}\bullet\{2,1\} & \rightarrow & \{2,2,1\}\label{eq:6py}
\end{eqnarray}

\begin{figure}[tph]
\textcolor{black}{\centering \includegraphics[bb=55bp 80bp 558bp 404bp,clip,scale=0.7]{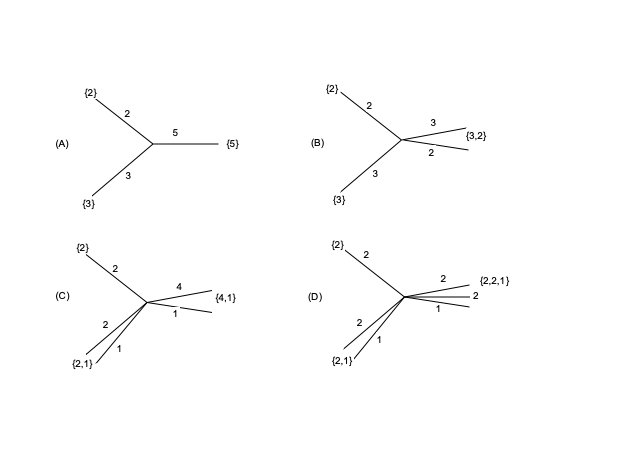}\caption{\label{fig:3-pt-int}{\footnotesize{}(A), (B), (C) and (D) illustrate
three-point, four-point, five-point and six-point interactions described
in \eqref{eq:3pt},\eqref{eq:4pt},\eqref{eq:5pt} and \eqref{eq:6py}
respectively.}}
}
\end{figure}

\eqref{eq:3pt}, \eqref{eq:4pt}, \eqref{eq:5pt} and \eqref{eq:6py}
are like three-point, four-point, five-point and six-point interaction
respectively as illustrated in figure \eqref{fig:3-pt-int}. This naive
way of combining $P\{2\}$ and $P\{3\}$ almost produces $P\{5\}$
except some over-counting. Given $P\{n\}$ or $P(n)$, although there
are formula which take close to $P\{n+1\}$ or $P(n+1)$, only way
to get the exact result is to use the definition of partition.\textcolor{black}{}
We will use the term \emph{interaction} for combination of parts and
\emph{merger} for combination of partition.

\textcolor{black}{When two separate systems merge into one, these
interactions take place at the microscopic level.} \textcolor{black}{We
wish to arrange the merger
\begin{eqnarray*}
P\{2\}\bullet P\{3\} & \rightarrow....\rightarrow P\{2,3,n\}\rightarrow...\rightarrow & P\{5\}
\end{eqnarray*}
into a series of intermediate steps labelled by }\textcolor{black}{\emph{distance}}\textcolor{black}{{}
$n$ such that for $P\{2,3,n=\infty\}=P\{2\}\bullet P\{3\}$ (two
systems are completely separate) and $P\{2,3,n=0\}=P\{5\}$ (systems
have completely merged). $n$ is a measure of oneness which defines
distance from complete merger. This distance emerging from interacting
system defines a manifold. This is one of the main themes of the project.
Interactions are fundamental from which spacetime emerges. This is
opposite of treating spacetime as given and interactions happening
in it. }All the above discussion can be summed up as the principle
of emergent causality. 

Two of the sharpest predictions of the model is on information paradox\citep{Harlow:2014yka,Giddings:1995gd,Mathur:2009hf}.
While holography\citep{Horowitz:2006ct}, soft theorems\citep{Hawking:2016msc,Haco:2018ske}
and other tools \citep{Susskind:2014rva,Faulkner:2013ica,Nishioka:2009un,Mathur:2008nj,Papadodimas:2012aq}
have shed light on the paradox and microstates of some black holes
have also been counted \citep{Strominger:1996sh,Sen:2014aja}, but
puzzle is still far from being solved. When two black holes merge,
gravitational radiation carry information. At least soft part of the
radiation is known to carry some information \citep{Chatterjee:2017zeb}.
At the same time, Hawking radiation is also expected to carry information.
This results in ambiguity. Does gravitational radiation and Hawking
radiation, which are of classical and quantum origin respectively,
both carry information, how much and what exactly is the information?
This model predicts that gravitational radiation carries black-hole
entropy worth of information. Secondly, we know that Hawking radiation
is a result of difference in vacuum due to curved background. Energy
conservation says that energy must drain out of black hole but the
mechanism is still not known \citep{Parikh:2004rh,Strominger:1994tn}.
This model suggests a clear picture of the mechanism. Another immediate
question might be how does this relate to holography and stringy microstates
of black hole? Please refer to the conclusion section for comments
on this issue.

In an upcoming up paper we will give two experimentally verifiable
predictions. Correction to Schwarzschild metric and gravitational
radiation. Once measured it will let us retrieve information released
during merger up-to $O(G^{3})$. 

\section{Outline}

\begin{tcolorbox}[colback=blue!5!white,colframe=blue!75!black]
Now we state the outline of the project. We postulate that black hole
of mass $M\in\mathbb{N}$ is described by $P\{M^{2},w\}$ (partition
of $M^{2}$ with weight $w$) as defined in section \eqref{sec:Partition}.
We also calculate asymptotes of such a partition. Merger of two black
holes of mass $M$ and $m$ to form a black hole of mass $M+m$ is
then given by merger of the partitions\textcolor{black}{
\begin{eqnarray}
P\{M^{2},w\}\bullet P\{m^{2},w\} & \rightarrow...\rightarrow P\{M^{2},m^{2},w,n\}\rightarrow...\rightarrow & P\{(M+m)^{2},w\}\label{eq:bhmerge1}
\end{eqnarray}
}

When the two black holes are far away ($n=\infty$) total degeneracy
is $P\{M^{2},w\}\bullet P\{m^{2},w\}$. Finally when the black holes
have merged, degeneracy is $P\{(M+m)^{2},w\}$ and $n=0$. Intermediate
state when the two black holes are at distance $n$ is denoted by
$P\{M^{2},m^{2},w,n\}$. This is the physical meaning of \eqref{eq:bhmerge1}. 
When separation $n$ between two black holes is much larger than their
Schwarzschild radius then the leading effect of merger is given by
the motion of one black hole in the background due to the other.
To find metric it is necessary to switch off the gravitational field
of one of them. That is taking test particle limit. It is
done by taking $m$ small enough such that $P\{m^{2},w\}\sim1$. The
process is then given by \textcolor{black}{
\begin{eqnarray*}
P\{M^{2},w\} & \rightarrow...\rightarrow P\{M^{2},m^{2},w,n\}\rightarrow...\rightarrow & P\{(M+m)^{2},w\}
\end{eqnarray*}
}Note that $P\{(M+m)^{2},w\}>>P\{M^{2},w\}$ even if $P\{m^{2},w\}\sim1$.
In the regime where $n>>GM>>Gm$ the leading effect of merger is then
given by geodesic motion of test black hole in the background due
to other black hole. We give prescription for the intermediate state
$P\{M^{2},m^{2},w,n\}$ in section \eqref{sec:Formation-of-Partition}.
From geodesic motion one can derive metric. We show that with this
prescription, the merger is identical to a particle falling in Schwarzschild
metric in section \eqref{sec:Emergent-Metric}.\\

In section \eqref{sec:Lorentz-transformation} we show that Lorentz
transformation is isomorphic to transformation under which local information
remain invariant. This is necessary to prove some of the assumptions
used in earlier sections. We find parts corresponding to the intermediate
states in section \eqref{sec:Observables}. These are the observables.
We conclude the paper with the results, experimental predictions and
future projects in section \eqref{sec:Conclusion}.\\

The reader might be wondering how come merger of degeneracy of black
holes describe interaction even when the separation is much larger
than their Schwarzschild radius. Usual picture in physics is the following.
Region inside the event horizon is believed to be black hole. That
is degeneracy of black hole are localized close to the singularity
or atmost delocalized upto the horizon. 
\end{tcolorbox}

\begin{tcolorbox}[colback=blue!5!white,colframe=blue!75!black]
This matter content inside horizon
then sources metric field according to Einstein equation. Metric
field then interacts with far away black hole. So only when black
holes are closer than the Schwarzschild radius that interaction can
possibly be described by the merger of degeneracy. By the end of section
\eqref{sec:Emergent-Metric} it becomes clear that the postulate is
only partially stated. Black hole and spacetime metric are not two
different objects. They are one and the same. They are unified and
replaced by the concept of partiton $P\{M^{2},w\}$. We call this
blackhole-space. \\

\textit{\small{}This is the limitation of presenting as postulate.
As the corresponding concept in physics is developed only after the
full understanding. However for the sake of some clarity this is approximate
chart of the project. }\textit{\textcolor{black}{\small{}These boxed
comments give physical intuitions. It is helpful to develop physical
intuition parallely. But if one feels uneasy then one may choose to
ignore the boxes in the first pass. At the end of section \eqref{sec:Emergent-Metric}
when the map to physics is eshtablished then one can come back and
read these comments.}}{\small\par}
\end{tcolorbox}

\section{Partition \label{sec:Partition}}

\textcolor{black}{We consider the following sequence $\{m^{2}:m\in\mathbb{N}\}=\{1,4,9,16,25,...\}$.
Define weighted partition $P\{m^{2},w\}$ with given weight $w\in\mathbb{{N}}$
as the following. Suppose a part is
\begin{eqnarray*}
m^{2} & = & \sum_{i}^{j}{\color{red}k_{i}}m_{i}^{2}
\end{eqnarray*}
then include $W=w^{\sum_{i}^{j}{\color{red}k_{i}}}$ copies of the
part in the partition. For example 
\begin{eqnarray*}
9 & = & {\color{green}{\color{red}1}}\times9={\color{red}2}\times4+{\color{red}1}\times1={\color{red}1}\times4+{\color{red}5}\times1={\color{red}9}\times1\\
P\{9,w\} & = & \left\{ \underset{w^{{\color{red}1}}}{\underbrace{\{9\},...,\{9\}}},\underset{w^{{\color{red}2+1}}}{\underbrace{\{4,4,1\},...,\{4,4,1\}}},\underset{w^{{\color{red}1+5}}}{\underbrace{\underset{5}{\left\{ 4,\underbrace{1,...,1}\right\} },...,\underset{5}{\left\{ 4,\underbrace{1,...,1}\right\} }}},\underset{w^{{\color{red}9}}}{\underbrace{\underset{9}{\left\{ \underbrace{1,...,1}\right\} },...,\underset{9}{\left\{ \underbrace{1,...,1}\right\} }}}\right\} \\
P(9,w) & = & w^{{\color{red}1}}+w^{{\color{red}2+1}}+w^{{\color{red}1+5}}+w^{{\color{red}9}}\\
P(9,w=10) & = & 10+10^{3}+10^{6}+10^{9}=1001001010
\end{eqnarray*}
Generating function of the above partition is
\begin{eqnarray}
Z & = & \prod_{m^{2}}\frac{1}{1-wz^{m^{2}}}\label{eq:generatsq}
\end{eqnarray}
First thing to study about a partition is its asymptotic expansion.
Asymptotic behavior is calculated as following.
\begin{eqnarray}
Z(z) & =\sum_{m\geq0}P(m)z^{m} & =\prod_{m\geq1}\frac{1}{1-wz^{m^{2}}}\label{eq:partition generating}
\end{eqnarray}
Steps of the proof follows \href{https://oeis.org/A006906/a006906.txt}{OEIS-A006906}.
Product is well-defined for $|z|<1$ except for poles at $z_{k^{2}}=w^{-1/k^{2}}\alpha_{k^{2}}$
where $\alpha_{k^{2}}$ are the $\left(k^{2}\right)^{th}$ roots of
unity. Coefficient $P(m^{2})$ of $z^{m^{2}}$ can be calculated by
taking contour integral around $z=0$.
\begin{eqnarray*}
P\left(m^{2}\right) & =\frac{1}{2\pi i}\oint\frac{Z(z)}{z^{m^{2}+1}} & =\frac{1}{2\pi i}\oint\prod_{n\geq1}\frac{dz}{\left(z^{m^{2}+1}\right)\left(1-wz^{n^{2}}\right)}
\end{eqnarray*}
Depending on the radius of contour, integral will pick up poles. Taking
circular contour, first set of poles appear at radius $|z|=w^{-1}$.
This will give the leading contribution to the contour integral. Next
set of poles appear at $|z|=w^{-1/4}$. 
\begin{eqnarray*}
P\left(m^{2}\right) & = & \frac{1}{2\pi i}\oint_{|Z|}\frac{Z(z)dz}{\left(z^{m^{2}+1}\right)}-R\left(w^{-1}\right)
\end{eqnarray*}
As we will see, residue $R(w^{-1})$ is of order $w^{m^{2}}$. Integrand
inside the contour is of order $w^{m^{2}/4}$. So the contour can
be anywhere between $w^{-1}<|z|<w^{-1/4}$ to get the leading contribution.
\begin{eqnarray}
\lim_{m\rightarrow\infty}P\left(m^{2}\right) & = & -R(w^{-1})\label{eq:lead residue}\\
 & = & -\lim_{z\rightarrow w^{-1}}\frac{\left(z-w^{-1}\right)}{z^{m^{2}+1}\left(1-wz\right)}\prod_{n\geq2}\frac{1}{1-wz^{n^{2}}}\nonumber \\
 & = & \lim_{z\rightarrow w^{-1}}\frac{1}{wz^{m^{2}+1}}\prod_{n\geq2}\frac{1}{1-wz^{n^{2}}}\nonumber \\
 & = & w^{m^{2}}\prod_{n\geq2}\frac{1}{1-w^{1-n^{2}}}\nonumber \\
 & = & r_{1}e^{G^{2}m^{2}}\nonumber 
\end{eqnarray}
where $r_{1}=\prod_{n\geq2}\frac{1}{1-w^{1-n^{2}}},G^{2}=\ln w$.
In this note we will assume $w=10$ whenever we want to get some estimate.
For $w=10,r_{1}=1.001001011$. This is the leading behavior or the
asymptotic behavior of the square partition. To calculate correction
to asymptotic behavior stretch the contour to $w^{-1/4}<|z|<w^{-1/9}$.
Contribution will come from next set of poles at $z_{0}=w^{-1/4}\{1,i,-1,-i\}$.
\begin{eqnarray*}
P\left(m^{2}\right) & = & -R(w^{-1})-R\left(w^{-1/4}\right)-R\left(iw^{-1/4}\right)-R\left(-w^{-1/4}\right)-R\left(-iw^{-1/4}\right)\\
R\left(z_{0}\right) & = & \lim_{z\rightarrow z_{0}}\frac{\left(z-z_{0}\right)}{z^{m^{2}+1}\left(1-wz^{4}\right)}\prod_{n\geq1,n\neq2}\frac{1}{1-wz^{n^{2}}}\\
 & = & \lim_{z\rightarrow z_{0}}\frac{\left(z-z_{0}\right)}{wz^{m^{2}+1}\left(z-z_{0}\right)\left(z-iz_{0}\right)\left(z+z_{0}\right)\left(z+iz_{0}\right)}\prod_{n\geq1,n\neq2}\frac{1}{1-wz^{n^{2}}}\\
 & = & -\frac{1}{wz_{0}^{m^{2}+4}\left(1-i\right)\left(1+1\right)\left(1+i\right)}\prod_{n\geq1,n\neq2}\frac{1}{1-wz_{0}^{n^{2}}}\\
 & = & -\frac{z_{0}^{-m^{2}}}{4}\prod_{n\geq1,n\neq2}\frac{1}{1-wz_{0}^{n^{2}}}\\
R\left(w^{-1/4}\right) & = & -\frac{w^{m^{2}/4}}{4}\prod_{n\geq1,n\neq2}\frac{1}{1-w^{1-n^{2}/4}}\\
R\left(iw^{-1/4}\right) & = & -i^{m^{2}}\frac{w^{m^{2}/4}}{4}\prod_{n\geq1,n\neq2}\frac{1}{1-i^{n^{2}}w^{1-n^{2}/4}}\\
R\left(-w^{-1/4}\right) & = & -(-1)^{m^{2}}\frac{w^{m^{2}/4}}{4}\prod_{n\geq1,n\neq2}\frac{1}{1-(-1)^{n^{2}}w^{1-n^{2}/4}}\\
R\left(-iw^{-1/4}\right) & = & -(-i)^{m^{2}}\frac{w^{m^{2}/4}}{4}\prod_{n\geq1,n\neq2}\frac{1}{1-(-i)^{n^{2}}w^{1-n^{2}/4}}
\end{eqnarray*}
If $m$ is even 
\begin{eqnarray*}
P\left(m^{2}\right)+R(w^{-1}) & = & \frac{w^{m^{2}/4}}{4}\\
 &  & \prod_{n\geq1,n\neq2}\left(\frac{1}{1-w^{1-n^{2}/4}}+\frac{1}{1-i^{n^{2}}w^{1-n^{2}/4}}+\frac{1}{1-(-1)^{n^{2}}w^{1-n^{2}/4}}+\frac{1}{1-(-i)^{n^{2}}w^{1-n^{2}/4}}\right)
\end{eqnarray*}
If $m$ is odd
\begin{eqnarray*}
P\left(m^{2}\right)+R(w^{-1}) & = & \frac{w^{m^{2}/4}}{4}\\
 &  & \prod_{n\geq1,n\neq2}\left(\frac{1}{1-w^{1-n^{2}/4}}+\frac{i}{1-i^{n^{2}}w^{1-n^{2}/4}}-\frac{1}{1-(-1)^{n^{2}}w^{1-n^{2}/4}}-\frac{i}{1-(-i)^{n^{2}}w^{1-n^{2}/4}}\right)
\end{eqnarray*}
Thus
\begin{eqnarray*}
P\left(m^{2}\right) & = & r_{1}e^{G^{2}m^{2}}+r_{2^{2}}e^{G^{2}m^{2}/4}+O(e^{G^{2}m^{2}/9})\\
r_{2^{2}}=r_{4} & = & \begin{cases}
\frac{1}{4}\prod_{n\geq1,n\neq2}\left(\frac{1}{1-w^{1-n^{2}/4}}+\frac{1}{1-i^{n^{2}}w^{1-n^{2}/4}}+\frac{1}{1-(-1)^{n^{2}}w^{1-n^{2}/4}}+\frac{1}{1-(-i)^{n^{2}}w^{1-n^{2}/4}}\right) & m=2k\\
\frac{1}{4}\prod_{n\geq1,n\neq2}\left(\frac{1}{1-w^{1-n^{2}/4}}+\frac{i}{1-i^{n^{2}}w^{1-n^{2}/4}}-\frac{1}{1-(-1)^{n^{2}}w^{1-n^{2}/4}}-\frac{i}{1-(-i)^{n^{2}}w^{1-n^{2}/4}}\right) & m=2k+1
\end{cases}
\end{eqnarray*}
For $w=10$
\begin{eqnarray}
r_{2^{2}}=r_{4} & = & \begin{cases}
-0.01112435883 & m=2k\\
-0.00626025134 & m=2k+1
\end{cases}\label{eq:r4}
\end{eqnarray}
Similarly one can find out higher corrections to $P\left(m^{2}\right)$.
Formally it will be
\begin{eqnarray*}
P\left(m^{2}\right) & = & \sum_{n\geq1}r_{n^{2}}e^{G^{2}m^{2}/n^{2}}\\
r_{w,n^{2}} & = & \frac{1}{n^{2}}\prod_{k\neq n}\sum_{j=1}^{n^{2}}\left(\frac{e^{2\pi j\frac{m^{2}}{n^{2}}i}}{1-e^{2\pi j\frac{k^{2}}{n^{2}}i}w^{1-\frac{k^{2}}{n^{2}}}}\right)
\end{eqnarray*}
We see partition is separated into asymptotes $P_{n}(m^{2})=r_{n^{2}}e^{G^{2}m^{2}/n^{2}}$
labelled by integer $n$. }

\begin{figure}[tph]
\textcolor{black}{\centering \includegraphics[scale=0.6]{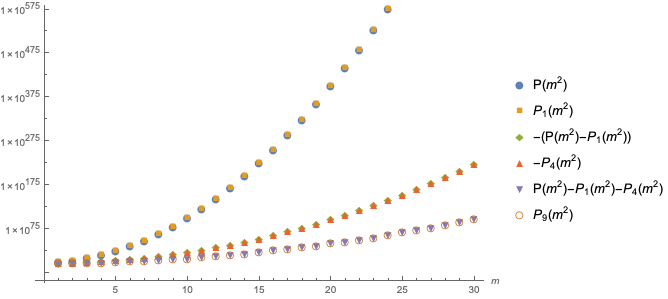}\caption{\textcolor{black}{\label{fig:sq asymptotes}}{\footnotesize{}Above
figure has six semilog plots related to partition of squares on natural
number }\textcolor{black}{\footnotesize{}$s(m)=\{1,4,9,...,m^{2},...:m\in\mathbb{{N}}\}$}{\footnotesize{}.
Plot in blue dots is partition of $s(m)$. Yellow squares represent
the plot of first asymptote calculated analytically. Yellow squares
and blue dots can be seen to lie almost on top of each other. Green
diamonds represent the difference between partition (blue dots)and
the first asymptote (yellow squares). Red triangles represent second
asymptote calculated analytically. Green diamonds and red triangles
can be seen to lie almost on top of each other. Violet inverted triangles
represent the difference between partition (blue dots)and the sum
of first (yellow squares) and second (red triangles) asymptotes. Yellow
circles represent third asymptote calculated analytically. Violet
inverted triangles and yellow circles can be seen to lie almost on
top of each other.}}
}
\end{figure}

\textcolor{black}{Leading asymptote $\sim e^{G^{2}m^{2}}$ is the
first motivation to consider this specific form of partition. Form
of $n^{th}$ subleading asymptote is $e^{G^{2}m^{2}/n^{2}}$. This
is shown in figure \eqref{fig:sq asymptotes}. Second motivation is
appearance of $n^{2}$ in the denominator in the exponent. Third motivation
is weight $G$ of the partition. Their physical interpretations will
become clear in the following sections. }

\textcolor{black}{There are two regimes. First one is the small $n$
regime where $n<Gm$. This regime is simple from partition perspective
and most visible in the graph. The coefficient $r_{1}\sim P(m^{2})e^{-G^{2}m}$.
On the other hand this regime is difficult because $e^{G^{2}m^{2}/n^{2}}>1$.
Hence perturbative analysis is not possible. Most of the degeneracy
is contained in this regime. Other is large $n$ regime $n>>Gm$.
This regime represent fine deviations of partition from the leading
asymptotic behavior and is difficult to see in the graph. As we will
see in section \eqref{sec:Observables}, physical meaning of the coefficients
$r_{n^{2}}$ become more and more convoluted as $n$ increases. On
the other hand, since $e^{G^{2}m^{2}/n^{2}}\sim1\implies\frac{\Delta P_{n}}{\Delta(Gm)}=r_{n^{2}}\left(\frac{2Gm}{n^{2}}\right)e^{G^{2}m^{2}/n^{2}}<<1$.
So perturbative analysis is possible. Our analysis in this paper will
mostly restrict to the large $n$ regime. These two regimes at the
two ends of the spectrum, lies at the heart of this work as we explain
in section \eqref{sec:Emergent-Metric}. Other consequences are described
in the conclusion \eqref{sec:Conclusion}.}

\textcolor{black}{}

\section{Partition Merger\label{sec:Formation-of-Partition}}

We now arrange the merger of two partitions \textcolor{black}{
\begin{eqnarray}
P\{M^{2}\}\bullet P\{m^{2}\} & \rightarrow...\rightarrow P\{M^{2},m^{2},n\}\rightarrow...\rightarrow & P\{(M+m)^{2}\}\label{eq:bhmerge2}
\end{eqnarray}
into a series of intermediate steps labelled by }\textcolor{black}{\emph{distance}}\textcolor{black}{{}
$n$ such that for $P\{M^{2},m^{2},n=\infty\}=P\{M^{2}\}\bullet P\{m^{2}\}$
(two systems are completely separate) and $P\{M^{2},m^{2},n=0\}=P\{(M+m)^{2}\}$
(systems have completely merged). Lets assume $M>>m$ and} $P(m^{2})\sim1$
(there is only one part), so that the process is\textcolor{black}{
\begin{eqnarray*}
P\{M^{2}\} & \rightarrow...\rightarrow P\{M^{2},m^{2},n\}\rightarrow...\rightarrow & P\{(M+m)^{2}\}
\end{eqnarray*}
We will call $P(m^{2})$ as test partition.}

\begin{tcolorbox}[colback=blue!5!white,colframe=blue!75!black]
\textcolor{black}{We postulate that black hole
of mass $M$ is represented by $P\{M^{2}\}$. So we will refer to
partition $P\{M^{2}\}$ as }\textcolor{black}{\emph{black hole}}\textcolor{black}{{}
interchangeably. Physical interpretation of \eqref{eq:bhmerge2} is
that black holes $P\{M^{2}\}$ and $P\{m^{2}\}$ merge to form $P\{(M+m)^{2}\}$}.
Physically $P(m^{2})\sim1$ approximation means turning off gravitational
field of $P\{m^{2}\}$. Reason will be clear by the end of the next
section. So the above process looks like\textcolor{black}{{} merger
of test black hole with a heavy black hole. We will refer to test
black holes as }\textcolor{black}{\emph{particle}}\textcolor{black}{{}
and use the phrase }\textcolor{black}{\emph{partitions merging, black
holes merging}}\textcolor{black}{{} and }\textcolor{black}{\emph{particle
falling}}\textcolor{black}{{} interchangeably. }
\end{tcolorbox} 

We proceed with the number of parts instead of explicit
parts. In section \eqref{sec:Observables} we will find out explicit
parts that constitute the intermediate states and claim them to be
observables\textcolor{black}{. So the process we study is the following
\begin{eqnarray}
P(M^{2}) & \rightarrow...\rightarrow P(M^{2},m^{2},n)\rightarrow...\rightarrow & P((M+m)^{2})\label{eq:formation}
\end{eqnarray}
Dynamics depend on the way we define $P(M^{2},m^{2},n)$. That will
also give physical meaning to $P(M^{2},m^{2},n)$ and $n$. In one
step change in partition is
\begin{eqnarray*}
... & \rightarrow P(M^{2},m^{2},n)\rightarrow P(M^{2},m^{2},n+\Delta n) & \rightarrow...
\end{eqnarray*}
which requires prescription for two things}
\begin{enumerate}
\item \textcolor{black}{Definition of $P(M^{2},m^{2},n)$.}
\item \textcolor{black}{Definition of $\Delta n\left(M^{2},m^{2},n\right)$.}
\end{enumerate}

\begin{tcolorbox}[colback=blue!5!white,colframe=blue!75!black]
\begin{itemize}
\item \textcolor{black}{Asymptotes are natural observables of partition.
When plotted on a graph, asymptotic behaviors are the most visible
thing. }
\item \textcolor{black}{Initial state $P(M^{2},m^{2},\infty)$ represents
a test particle far from black hole and the final state $P(M^{2},m^{2},0)$
is when the particle has fallen in it. Velocity of particle at intermediate
state $P(M^{2},m^{2},n)$ is also observable when it is at finite
distance from the black hole.}
\end{itemize}
\textcolor{black}{We want to associate these two observables. This
motivates us to associate intermediate states with the asymptotes. }
\end{tcolorbox}

\textcolor{black}{Define $\Delta t\left(M^{2},m^{2},n\right)$, as
change in $P_{n}$ relative to change in $P_{n}$ when $M\rightarrow M+1$.
\begin{eqnarray}
\Delta t\left(M^{2},m^{2},n\right) & \equiv & G\left(\frac{P_{n}((M+m)^{2})-P_{n}(M^{2})}{P_{n}((M+1)^{2})-P_{n}(M^{2})}\right)\label{eq:time}
\end{eqnarray}
For $n>>m$ 
\begin{eqnarray*}
\Delta t & = & Gm
\end{eqnarray*}
Define $P(M^{2},m^{2},n)$ and $\Delta n\left(M,m^{2},n\right)$ to
be
\begin{eqnarray}
P(M^{2},m^{2},n) & \equiv & \sum_{k\geq1}^{n-1}P_{k}(M^{2})+\sum_{k\geq n}^{\infty}P_{k}((M+m)^{2})\label{eq:intermediate}\\
\Delta v & \equiv & -\left(\frac{P_{n}((M+m)^{2})-P_{n}(M^{2})}{P_{n}(M^{2})}\right)\nonumber \\
\Delta n & \equiv & v\Delta t\label{eq:displacement}
\end{eqnarray}
As an intermediate step we have defined $v$. From now on discussions will be for $n>>m$. From
above prescription, when test partition is at separation $n$, degeneracy
of $n^{th}$ asymptote changes
\begin{eqnarray}
\Delta P_{n} & = & P_{n}((M+m)^{2})-P_{n}(M^{2})\nonumber \\
 & = & r_{n^{2}}e^{G^{2}(M+m)^{2}/n^{2}}-r_{n^{2}}e^{G^{2}M^{2}/n^{2}}\nonumber \\
 & = & r_{n^{2}}\left(\frac{2G^{2}Mm}{n^{2}}\right)e^{G^{2}M^{2}/n^{2}}\label{eq:info change}
\end{eqnarray}
and correspondingly
\begin{eqnarray}
\Delta v & = & -\left(\frac{\Delta P_{n}}{P_{n}}\right)\nonumber \\
 & = & -\frac{2G^{2}Mm}{n^{2}}\nonumber \\
 & = & -\frac{2GM}{n^{2}}\Delta t\label{eq:velocity}
\end{eqnarray}
With the condition $v(n=\infty)=0$ we get 
\begin{eqnarray*}
v(n) & = & -\sqrt{\frac{2GM}{n}}
\end{eqnarray*}
}

\textcolor{black}{Let $M^{2}=\sum_{i}k_{i}m_{i}^{2}$ be a state which
contributes to
\begin{eqnarray*}
\Delta P_{n} & = & \left(\frac{2G^{2}Mm}{n^{2}}\right)r_{n^{2}}e^{G^{2}M^{2}/n^{2}}
\end{eqnarray*}
Before merger, states $\{m_{i}^{2}\}$ are observable and will be
called }\textcolor{black}{\emph{space}}\textcolor{black}{{} as they
support the distance between the two partitions. }

\textcolor{black}{Upon merger, distance decreases and space is no
more observable. Information $\{m_{i}^{2}\}$ is released, states
$\{m_{i}^{2}\}$ become non-observable parts and degeneracy increases
by $w^{\sum_{i}k_{i}}$.
\begin{eqnarray*}
P(M^{2},m^{2},n) & \rightarrow & P(M^{2},m^{2},n)+w^{\sum_{i}k_{i}}
\end{eqnarray*}
In short, information associated with space is released and
space is converted into parts.
\begin{eqnarray*}
\text{Observable space} & \xrightarrow[\text{Information released}]{\text{Merger, Degeneracy increases}} & \text{Non-observable parts}
\end{eqnarray*}
In other words space at radius
$n$ holds
\begin{eqnarray*}
\Delta P_{n} & = & \left(\frac{2G^{2}Mm}{n^{2}}\right)r_{n^{2}}e^{G^{2}M^{2}/n^{2}}
\end{eqnarray*}
amount of information. Now imagine the reverse process of sending information into $P(M^{2})$.
Relative decrease of degeneracy will be 
\begin{eqnarray*}
\left(\frac{\Delta P_{n}}{P_{n}}\right) & = & -\frac{2G^{2}Mm}{n^{2}}
\end{eqnarray*}
As a result number of parts decrease by $\left(\frac{2G^{2}Mm}{n^{2}}\right)r_{n^{2}}e^{G^{2}M^{2}/n^{2}}$
which must be compensated by increase in space. Thus we find that
sending information into the system forces partitions to de-merge.
We will call this }\textcolor{black}{\emph{information pressure.}}\textcolor{black}{{}
As we discuss in conclusion, this may play an important role in Hawking
radiation.}

\textcolor{black}{With this prescription, we can give physical meaning
to the above process. If we identify $n$ with radius and $t$ with
time, then equation \eqref{eq:displacement} and \eqref{eq:velocity}
gives decrease in radius and velocity respectively. $v$ in a sense measures
rate of merger With $\Delta P_{n}$
change in degeneracy, $2G^{2}Mm$ volume of space is annihilated over
surface area of sphere and velocity reduces by $\frac{2G^{2}Mm}{n^{2}}$.
Effectively it appears like $P\{M^{2}\}$ annihilates space of volume
$2GM$ per unit time. $\frac{1}{n^{2}}$ factor in $\Delta v=\frac{2G^{2}Mm}{n^{2}}$
would give an impression that there is conservation and continuity
relation and space is flowing towards the partition. As a result,
space shrinks with velocity $v=\sqrt{\frac{2GM}{n}}$ and test partition
drags along with the space fabric.  }

\textcolor{black}{This would mean the system is causal as then the
information released at separation $n$ and $(n+1)$ would be related.
However, presence of $r_{n^{2}}$ in
\begin{eqnarray*}
P_{n} & = & r_{n^{2}}e^{G^{2}M^{2}/n^{2}}
\end{eqnarray*}
invalidates this effective picture. As we show in section \eqref{sec:Observables},
coefficients $r_{n^{2}}$ are nothing but partitions in convoluted
form. There is no exact relation between $r_{n^{2}}$ and $r_{(n+1)^{2}}$.
In other words, information at separation $n$ and $(n+1)$ is not
related. So there is no flow of space from radius $n+1$ to $n$.
Appearance of flow is only emergent. Information released at two consecutive
time slices are not related. This is how causality emerges. }

\begin{tcolorbox}[colback=blue!5!white,colframe=blue!75!black]
\textcolor{black}{Effectively it appears like the black hole of mass
$M$ annihilates space of volume $2GM$ per unit time. $\frac{1}{n^{2}}$
factor in $\Delta v=\frac{2G^{2}Mm}{n^{2}}$ would give an impression
that there is conservation and continuity relation and space is flowing
towards black hole and finally draining into it. As a result,
space shrinks with velocity $v=\sqrt{\frac{2GM}{n}}$ and test particle
drags along with the space fabric. }
\end{tcolorbox}

\begin{figure}[tph]
\textcolor{black}{\centering \includegraphics[scale=0.5]{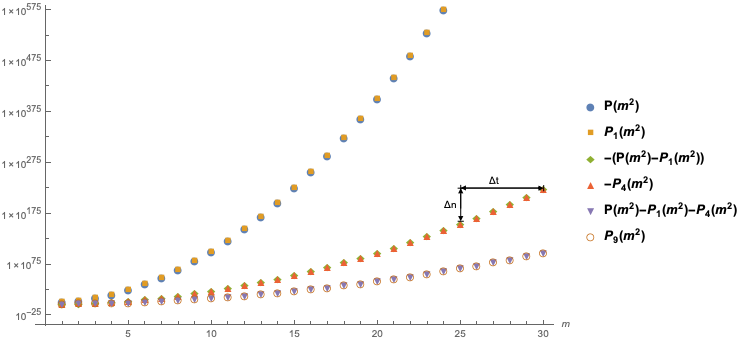}\caption{\textcolor{black}{\label{fig:space-time}}{\footnotesize{}Above figure
is same as figure \ref{fig:sq asymptotes} with additional details
which shows that time emerges from difference between initial and
final $m$ values and space emerges from difference in initial and
final values of asymptotes.}}
}
\end{figure}

\textcolor{black}{We see that time and space emerge from merger of
partition. Time and space is related to the shift in the x-axis and
y-axis respectively in the partition as shown in figure \eqref{fig:space-time}.
There is no meaning of space or time without merger.}

\textcolor{black}{}
\textcolor{black}{One can ask, why to start with the sequence $\{m^{2}:m\geq1\}=\{1,4,9,16,25,...\}$?
One motivation was to get a degeneracy which grows like $\sim e^{G^{2}m^{2}}$.
Now we have another completely different motivation. Suppose one had
started with $\{s(m):m\geq1\}$ where $s(m)$ is any increasing function
of $m$. All the above derivation will go through with the replacements
$m^{2}\rightarrow s(m),n^{2}\rightarrow s(n)$. We will end up with
$\Delta n=-\frac{G^{2}s'(M)}{s(n)}\Delta M$. However we saw that
interpreting $s(n)$ as surface area of sphere gives a nice picture.
This is the second motivation which connects $n^{th}$ asymptote to
sphere of radius $n$. }

\begin{figure}[tph]
\textcolor{black}{\centering \includegraphics[scale=0.6]{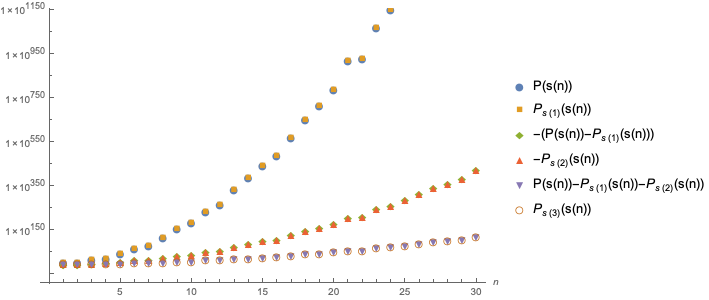}\caption{\textcolor{black}{\label{fig:sph asymptotes}}{\footnotesize{}Above
figure has six semilog plots related to partition of sequence }\textcolor{black}{\footnotesize{}$s=\{6,26,90,134,258,...\}$
where}{\footnotesize{} }\textcolor{black}{\footnotesize{}$s(n)$ is
number of integer solutions of $n\protect\geq\sqrt{x^{2}+y^{2}+z^{2}}>n-1$}{\footnotesize{}.
Plot in blue dots is partition of $s(n)$. Yellow squares represent
the plot of first asymptote calculated analytically. They can be seen
to lie almost on top of each other. Green diamonds represent the difference
between partition (blue dots)and the first asymptote (yellow squares).
Red triangles represent second asymptote calculated analytically.
Green diamonds and red triangles can be seen to lie almost on top
of each other. Violet inverted triangles represent the difference
between partition (blue dots)and the sum of first (yellow squares)
and second (red triangles) asymptotes. Yellow circles represent third
asymptote calculated analytically. Violet inverted triangles and yellow
circles can be seen to lie almost on top of each other.}}
}
\end{figure}

\textcolor{black}{So we choose $s(n)$ to be number of integer lattice
points between $n\geq r>n-1$. In other words, number of integer solutions
of $n\geq\sqrt{x^{2}+y^{2}+z^{2}}>n-1$. $s=\{6,26,90,134,258,...\}$
. For $n>>1,s(n)\rightarrow4\pi n^{2}$ which is surface area of sphere
of radius $n$. This also reproduces $\Delta v=-\frac{2G^{2}Mm}{n^{2}}$
for $M,n>>1$. First few asymptotes of partition are shown in figure
\eqref{fig:sph asymptotes}. For example, partition of $s(4)$ is
\begin{eqnarray*}
 &  & P{s(4),w=1}\\
 & = & \Bigg\{\left\{ 134\right\} ,\left\{ 90,26,6,6,6\right\} ,\left\{ \underset{4}{\underbrace{26,...,26}},\underset{5}{\underbrace{6,...,6}}\right\} ,\left\{ 26,\underset{18}{\underbrace{6,...,6}}\right\} \Bigg\}\\
 & = & \Bigg\{\left\{ s(4)\right\} ,\left\{ s(3),s(2),s(1),s(1),s(1)\right\} ,\left\{ \underset{4}{\underbrace{s(2),...,s(2)}},\underset{5}{\underbrace{s(1),...,s(1)}}\right\} ,\left\{ s(2),\underset{18}{\underbrace{s(1),...,s(1)}}\right\} \Bigg\}
\end{eqnarray*}
}

For all the calculation we will use $s(n)=n^{2}$. Many properties
are easy to illustrate using this sequence.\textcolor{black}{}

\textcolor{black}{Redoing the above calculation with $s(n)$ gives
the following result
\begin{eqnarray*}
P(s(m)) & = & \sum_{n\geq1}r_{s(n)}e^{G^{2}s(m)/s(n)}
\end{eqnarray*}
whose generating function is
\begin{eqnarray*}
Z(z) & =\sum_{n\geq0}P(n)z^{n} & =\prod_{n\geq1}\frac{1}{1-wz^{s(n)}}
\end{eqnarray*}
and 
\begin{eqnarray}
r_{w,s(n)} & = & \frac{1}{s(n)}\prod_{k\neq n}\sum_{j=1}^{s(n)}\left(\frac{e^{2\pi j\frac{s(m)}{s(n)}i}}{1-e^{2\pi j\frac{s(k)}{s(n)}i}w^{1-\frac{s(k)}{s(n)}}}\right)\label{eq:residue}
\end{eqnarray}
}

Given the sequence of integer \textcolor{black}{$s=\{6,26,90,134,258,...\}$,
all that we have done is to study partition of these numbers. At best
this contains radial evolution. How will the angular degree of freedom
emerge? Given the infinite sequence $s=\{6,26,90,134,258,...\}$,
$s$ is automatically mapped to solutions of $n\geq\sqrt{x^{2}+y^{2}+z^{2}}>n-1$.
Thus angular information is hidden in the pattern of the infinite
sequence. }

\begin{tcolorbox}[colback=blue!5!white,colframe=blue!75!black]
\textcolor{black}{Also, this is not surprising because angular variables
of a single black hole are not observable due to spherical symmetry.
To derive metric we will study an infalling particle in sections \eqref{sec:Emergent-Metric}
and \eqref{sec:Lorentz-transformation}. There angular dependence will
play a crucial role.}
\end{tcolorbox}

\section{Emergent Metric\label{sec:Emergent-Metric} }

\textcolor{black}{Discussion of the previous section has set up the
stage to derive the metric emerging from partition. In unit time when
$P_{n}(s(M))$ changes to $P_{n}(s(M+1))$, effectively it appears
like space shrinks with velocity $v=\sqrt{\frac{2GM}{n}}$ and test
partition drags along with the space fabric. Goal is to show that
the merger of test partition is identical to that of a fall into a
black hole. There is some literature on similar issue \citep{Czerniawski:2006sc}.}

\begin{tcolorbox}[colback=blue!5!white,colframe=blue!75!black]
\textcolor{black}{When separation between two black holes is much
larger than their Schwarzschild radius then the leading effect of
merger is given by the motion of test black hole in the background
due to other black hole. We now derive metric from the motion of test
particle. }
\end{tcolorbox}

\textcolor{black}{We start by distinguishing two frames}
\begin{enumerate}
\item \textcolor{black}{\emph{Schwarzschild frame}}\textcolor{black}{. Coordinates
are $(t_{s},n_{s},\Omega)$. $t_{s}$ is the time coordinate, $n_{s}$
is the radial coordinate which measures distance from centre of the
partition and $\Omega$ is the angular coordinate. This frame coincides
with the rest frame of an observer at $n_{s}=\infty$.  }
\item \textcolor{black}{Rest frame of the test partition hovering at radial distance
$n_{s}$. To stay at fixed $n_{s}$, it has to continuously
\textit{boost}. Coordinates are $(t_{h},n_{h},\Omega)$ which are functions
of $t_{s},n_{s}$. We will call it }\textcolor{black}{\emph{hovering
frame}}\textcolor{black}{.}
\end{enumerate}

\begin{tcolorbox}[colback=blue!5!white,colframe=blue!75!black]
\textcolor{black}{In the above definition of frames we can replace the word partition by black hole to get physical sense.\\
To realize hovering frame we have to boost a particle.
Hovering frame is boosted with respect to Schwarzschild frame. For
that first we need to understand energy-momentum and boost. So far
in our discussion we only had mass. When a particle falls in gravitational
field the leading effect is same as boost. This opens a possibility
to define boost in terms of merger of black holes. During merger of
partition, we will show that, deformation of asymptotes are isomorphic
to Lorentz transformation. This will define energy-momentum. }
\end{tcolorbox}

\textcolor{black}{In our derivation below, we will make number of
assumptions which we will prove in section \ref{sec:Lorentz-transformation}.
Metric at radial coordinate $n_{s}$ is nothing but local line element
in hovering frame. Our first assumption is that boost is well defined
so that hovering observer exists and line element in hovering frame
is given by
\begin{eqnarray}
\Delta s^{2} & = & -\Delta t_{h}^{2}+\Delta n_{h}^{2}\label{eq:linehov}
\end{eqnarray}
Non-trivial part is $-\Delta t_{h}^{2}$. $\Delta n_{h}^{2}$ follows
from the choice of $s(n)$. We want to express the line element in
Schwarzschild coordinates. Consider two events at $n_{s}$ ($\Delta n_{s}=0$)
and separated by unit step in Schwarzschild frame. $\Delta n_{s}=0$
implies that the events are at rest in hovering frame and are boosted
with respect to Schwarzschild frame. Our second assumption is that
time separation in Schwarzschild frame between the events is
\begin{eqnarray}
\Delta t_{s} & =GE= & Gm\cosh u\label{eq:boost-t}
\end{eqnarray}
where $u=\tanh^{-1}\left(v\right),v=\sqrt{\frac{2GM}{n}}$. Information
emitted in a unit step is the fundamental event and physically observable.
So we assume that unit step is frame invariant. Events separated by
unit step in Schwarzschild frame will also be separated by unit step
in hovering frame. In hovering frame by definition $\Delta t_{h}=Gm,\Delta n_{h}=0$.
So $\Delta t_{s},\Delta t_{h}$ give time interval between same two
events
\begin{eqnarray}
\Delta t_{h} & = & \frac{\Delta t_{s}}{\cosh u}=\sqrt{1-v^{2}}\Delta t_{s}\label{eq:t-t'}
\end{eqnarray}
Consider a rod of rest length $L_{h}=L$. To stay at fixed $n_{s}$
it has to hover over $n_{s}$ and hence at rest in hovering frame.
Using the above relation between time intervals and equation \eqref{eq:linehov},
relation between length of rod in Schwarzschild frame and hovering
frame is
\begin{eqnarray}
L_{h} & = & \frac{L_{S}}{\sqrt{1-v^{2}}}\nonumber \\
\implies\Delta n_{h} & = & \frac{\Delta n_{s}}{\sqrt{1-v^{2}}}\label{eq:n-n'}
\end{eqnarray}
Equations \eqref{eq:t-t'} and \eqref{eq:n-n'} give the relation
between hovering coordinates and Schwarzschild coordinates. Substituting
them in equation \eqref{eq:linehov} we get
\begin{eqnarray}
\Delta s^{2} & = & -\left(1-v^{2}\right)\Delta t_{s}^{2}+\frac{\Delta n_{s}^{2}}{1-v^{2}}+d\Omega^{2}\nonumber \\
 & = & -\left(1-\frac{2GM}{n}\right)\Delta t_{s}^{2}+\left(1-\frac{2GM}{n}\right)^{-1}\Delta n_{s}^{2}+d\Omega^{2}\label{eq:metric}
\end{eqnarray}
Lorentz invariance of unit step is a crucial input which allows to
relate events in various frames. }

\begin{figure}[tph]
\textcolor{black}{\centering \includegraphics[bb=0bp 85bp 394bp 306bp,clip,scale=0.7]{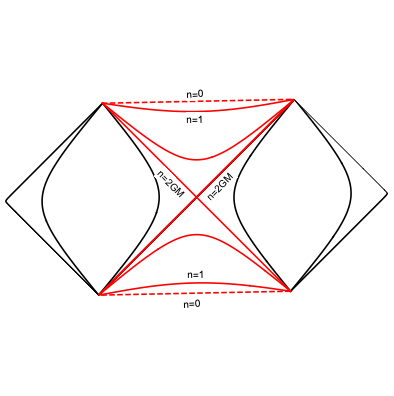}}

\textcolor{black}{\caption{\textcolor{black}{\label{fig:blackhole}}{\footnotesize{}Penrose diagram
of Schwarzschild black hole of mass $M$. $n$ is Schwarzschild radius.
$n=0$ is singularity and $n=2GM$ is the horizon. Degeneracy of black
hole is believed to be localized within the horizon (mostly near singularity).
It is denoted by red constant $n$ slices. Information is localized
within horizon. Metric sourced by the black hole extends upto asymptotic
regions. This is depicted by black constant $n$ slices.}}
}
\end{figure}

\begin{figure}[tph]
\textcolor{black}{\centering \includegraphics[bb=0bp 90bp 385bp 302bp,clip,scale=0.7]{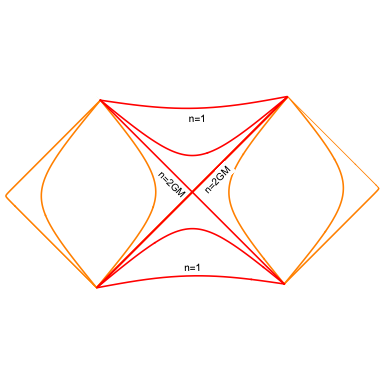}}

\textcolor{black}{\caption{\textcolor{black}{\label{fig:blackhole-space}}{\footnotesize{}Penrose
diagram of blackhole-space or partition $P\{M^{2}\}$. Partition ends
at $n=1$ as $1$ is the smallest natural number. There is no $n=0$
region. Most of the degeneracy is localized within the small $n$
regime $n<GM$. That is depicted by red constant $n$ slices. This
effectively appears like black hole with event horizon. Orange constant
$n$ slices in the large $n$ regime $n>GM$ depict that degeneracy
is present in this region also. Degeneracy is distributed over all
$n$. There is no spacetime metric for a single partition. Metric emerges
only in the context of merger of two partitions. Separate notion of
black hole (degeneracy of black hole) and spacetime metric is unified
and replaced by the concept of partition or blackhole-space. This
whole thing is one fundamental object $P\{M^{2}\}$ that can be measured
as discussed in conclusion \eqref{sec:Conclusion}.}}
}
\end{figure}

\textcolor{black}{This is Schwarzschild metric of a black hole of
mass $M$ and gravitational constant $G=\sqrt{\ln w}$. Secondly,
horizon corresponds to $2GM=n$. There is no $n=0$ region as $n=1$
is the smallest natural number. Going back to partition
\begin{eqnarray*}
P\left(M^{2}\right) & = & \sum_{n\geq1}r_{n^{2}}e^{G^{2}M^{2}/n^{2}}
\end{eqnarray*}
region close and far from origin is described by small $n$ regime
and large $n$ regime of the partition respectively. Sharp notion
of horizon enclosing all the information of black hole has disappeared.
Partition is not localized within the horizon but extends all the
way to the asymptotic region. There is no metric for a single partition.
Metric emerges only in the context of merger of two partitions. Separate
notion of matter (or black hole localized within horizon) and space-time
metric (empty space far away from origin) is unified by partition.
Effectively it appears as if the black hole has blurred out all over
the space. We will call it }\textcolor{black}{\emph{blackhole-space.
}}\textcolor{black}{Notion of black hole localized in spacetime and
curving it, is unified and replaced by one fundamental object blackhole-space
as depicted in figure \eqref{fig:blackhole} and \eqref{fig:blackhole-space}.
This opens up the possibility of measuring the whole partition in
one go as we discuss in conclusion. This establishes the following
result}\\

\begin{tcolorbox}[colback=blue!0!white,colframe=blue!75!black]
\begin{center}
\textcolor{black}{$P\{M^{2}\}=\text{Blackhole-space}$}
\par\end{center}
\end{tcolorbox}

So far all the discussion is with natural numbers, where as in physics we deal with unitful quantities. To get unitless quantity we divide by Planck unit. For example, for $M=1kg$. Since $GM$ has units of length, it has to be divided by $l_{p}$. Leading degeneracy at $n=1$ grows like
\begin{eqnarray*}
P_{1}\left(M^{2}\right) =  r_{1}e^{G^{2}/l_{p}^{2}} = r_{1}e^{Gc^{3}/\hbar} 
\end{eqnarray*}
where $l_{p}$ is Planck length. Degeneracy depends on $\hbar$ since  $n$ is just a natural number independent of $\hbar$. Unitless radius of event horizon is given by $n_{h}=2GM/l_{p}=2\sqrt{\frac{Gc^3}{\hbar}}$. Degeneracy corresponding to $n^{th}_h$ asymptote is
\begin{eqnarray*}
P_{n_{h}}\left(M^{2}\right) = r_{n_{h}}e^{G^{2}M^{2}/n_{h}^{2}}= r_{n_{h}}e^{1/4} 
\end{eqnarray*}
$\hbar$ dependence in  $n_{h}$ and $GM$ cancels out. So degeneracy on the surface of event horizon is independent of $\hbar$ as expected. 

\textcolor{black}{Now the reader may kindly read the previous boxed
comments to get the physical intuition. }

\textcolor{black}{}
\textcolor{black}{}
\textcolor{black}{In the above derivation we have used three assumptions:
equation \eqref{eq:linehov}, \eqref{eq:boost-t} and that unit step
is frame invariant. In the next section we will prove these assumptions. }

\section{Lorentz transformation\label{sec:Lorentz-transformation}}

In the previous section we found Schwarzschild metric. One of the
assumptions was energy-momentum relation. We show that as the test
particle merges with black hole, the leading effect is a boost. So
that by studying effect of merger on rest mass we will derive energy-momentum
relation. 

Leading effect of gravity is captured by equivalence principle, according
to which gravity is identical to acceleration locally. Motion can
be locally approximated by inertial frames and there are no local
experiments which can detect gravity. In this model, angular distribution
of lattice points is the local information. Taking hint from gravity,
we ask the following question. What are the transformations under
which this information remain invariant? In other words, what are
the transformations under which angular part of line element on a
sphere is invariant? We show that these transformations are isomorphic
to Lorentz transformation. \textcolor{black}{That will define energy-momentum
relation \eqref{eq:boost-t} and line element \eqref{eq:linehov}. }

Line element on a sphere is given by
\begin{eqnarray*}
dl^{2} & = & n^{2}d\theta^{2}+n^{2}\sin^{2}\theta d\phi^{2}
\end{eqnarray*}
Consider the transformation
\begin{eqnarray*}
n & = & K(\theta')n'+O(n'^{-p})\\
\theta & = & g(\theta')n'+g^{0}(\theta')+O(n'^{-p-1});p\geq0\\
\phi & = & \phi'
\end{eqnarray*}
Throughout the discussion we will assume azimuthal symmetry. Invariance
of the metric component $g_{\theta\theta}=n^{2}$ gives
\begin{eqnarray*}
g_{\theta'\theta'} & = & K(\theta')^{2}n'^{2}\left(\frac{dg(\theta')}{d\theta'}n'^{2}+\frac{dg^{0}(\theta')}{d\theta}\right)+O(n')=n'^{2}+O(n')\\
\implies g(\theta') & = & c\\
\frac{1}{k(\theta')} & = & \frac{dg^{0}(\theta')}{d\theta}
\end{eqnarray*}
Invariance of the metric component $g_{\phi\phi}=n^{2}\sin^{2}\theta$
gives
\begin{eqnarray*}
g_{\phi'\phi'} & = & k(\theta')^{2}n'^{2}\sin^{2}g^{0}+O(n')=n'^{2}\sin^{2}\theta'+O(n')\\
\implies\frac{dg^{0}(\theta')}{d\theta} & = & \frac{\sin g^{0}}{\sin\theta'}
\end{eqnarray*}
Choosing $c=0$ we get 
\begin{eqnarray}
n & = & K(\theta')n'\label{surface trans1}\\
\tan\left(\frac{g^{0}}{2}\right) & = & e^{-\nu}\tan\left(\frac{\theta'}{2}\right)\label{eq:surface trans2}\\
K(\theta') & = & \cosh\nu+\cos\theta'\sinh\nu\nonumber 
\end{eqnarray}
where $\nu$ is some constant of integration. We will call the above
transformations as \emph{surface transformations}. 

To see the transformation of mass $m$ of test particle,\textcolor{black}{{}
the appropriate question is, how is the field of test black hole modified
under above transformation?} Space around test black hole of mass
$m$ shrinks by $\sqrt{\frac{2Gm}{n}}$ in unit time. Hence drag velocity
is 
\begin{eqnarray*}
v^{n} & = & \sqrt{\frac{2Gm}{n}}\\
v^{\theta} & = & v^{\phi}=0
\end{eqnarray*}
Under the above transformation
\begin{eqnarray}
v^{n'} & = & \frac{\partial n'}{\partial n}v^{n}=\sqrt{\frac{2G}{n'}\frac{m}{K^{3}}}\label{eq:vel trans}\\
v^{\theta'} & = & \frac{\partial\theta'}{\partial n}v^{n}=O\left(n'^{-3}\right)\nonumber \\
v^{\phi'} & = & 0\nonumber 
\end{eqnarray}
This shows that $m$ transforms like
\begin{eqnarray*}
m'(\theta') & = & \frac{m}{\left(\cosh\nu+\cos\theta'\sinh\nu\right)^{3}}
\end{eqnarray*}
Let us define
\begin{eqnarray*}
E\equiv m'_{00} & \equiv & \int m'Y_{00}d\Omega=m\cosh u\\
p_{x}\equiv m'_{10} & \equiv & \int m'Y_{10}d\Omega=m\sinh u\\
p_{y}\equiv m'_{1,-1} & \equiv & \int m'Y_{1,-1}d\Omega=0\\
p_{z}\equiv m'_{11} & \equiv & \int m'Y_{11}d\Omega=0
\end{eqnarray*}
One can see that
\begin{equation}
E^{2}-p_{x}^{2}-p_{y}^{2}-p_{z}^{2}=m^{2}\label{eq:energy-mom}
\end{equation}
is invariant under above transformation. 

When a test particle of mass $m$ merges with a black hole, time elapsed
in unit step is $Gm$, which is same as flux of space volume annihilated
by the test particle in unit time, $\frac{1}{8\pi}\int\frac{dv^{n}}{dt}\sin\theta d\theta d\phi ndn$.
This allows us to give second definition of $\Delta t$
\begin{eqnarray*}
\Delta t & \equiv & \frac{1}{8\pi}\int\frac{dv^{n}}{dt}\sin\theta d\theta d\phi ndn
\end{eqnarray*}

For a Lorentz transformed test particle, flux is
\begin{eqnarray*}
\frac{1}{8\pi}\int\frac{dv^{n'}}{dt}\sin\theta d\theta d\phi n'drn' & = & GE
\end{eqnarray*}
Thus when boosted test particle merges with black hole, time elapsed
in unit step is $GE=Gm\cosh u$. We will call $E$ as \emph{energy.}

Non-vanishing higher spherical harmonics indicate that space is not
annihilated isotropically. 
\begin{eqnarray*}
m'(\theta') & = & E+p_{x}Y_{10}+\sum_{l>1}m'_{lm}Y_{lm}
\end{eqnarray*}
For example, consider harmonics $l=0,1$. Space annihilated at $\theta=\pi$
is $E-p_{x}$ and at $\theta=0$ is $E+p_{x}$. Thus $p_{x}$ represents shift
in position. We will call it \emph{momentum}.

This motivates us to identify $Gm'_{00}$ as the time element $\Delta t$
and $Gm'_{10}$ as the length element $\Delta x$. So shift
per unit time is $\frac{\Delta x}{\Delta t}=\tanh u$. For hovering frame, equating shift to
drag $\sqrt{\frac{2GM}{\bar{n}}}$ gives $u=\tanh^{-1}\sqrt{\frac{2GM}{\bar{n}}}$
($\bar{n}$ is radial coordinate of heavy black hole). This proves
equation \eqref{eq:boost-t}. Identification along with equation \eqref{eq:energy-mom}
allows us to construct invariant length
\begin{eqnarray}
ds^{2} & = & -dt^{2}+dn^{2}+n^{2}d\theta^{2}+n^{2}\sin^{2}\theta d\phi^{2}\label{eq:inv length}
\end{eqnarray}
This completes the derivation of equation \eqref{eq:linehov}. 

In addition to invariance of angular part of line element on sphere,
we have to also satisfy invariance of $ds^{2}$.
\begin{eqnarray*}
t & = & a(t',\theta')n'+a^{0}(t',\theta')+O(n'^{-p-1})\\
n & = & K(\theta')n'+\rho(t'-n',\theta')+O(n'^{-p-1})\\
\theta & = & 2\arctan\left(e^{-\nu}\tan\frac{\theta'}{2}\right)+O(n'^{-p-1});p\geq0\\
\phi & = & \phi'
\end{eqnarray*}
Holding $t'-n'$ constant and for large $n'$, above ansatz of $n$
transformation is consistent with \eqref{surface trans1} and \eqref{eq:surface trans2}.
\begin{eqnarray*}
g^{n'n'} & = & \left(K-\frac{d\rho}{d(t'-n')}\right)^{2}-a^{2}=1\\
g^{t't'} & = & -\left(\frac{da}{dt'}n'+\frac{da^{0}}{dt'}\right)^{2}+\left(\frac{d\rho}{d(t'-n')}\right)^{2}=-1
\end{eqnarray*}
From second equation, comparing coefficients of $n'^{2}$ gives $a=a(\theta')$.
Substituting this in first equation gives
\begin{eqnarray*}
\rho & = & (t'-n')\rho(\theta')+c_{1}
\end{eqnarray*}
which substituting in second equation gives
\begin{eqnarray*}
a^{0} & = & t'a^{0}(\theta')+c_{2}
\end{eqnarray*}
So we get
\begin{eqnarray}
\left(K-\rho(\theta')\right)^{2}-a^{2} & = & 1\label{eq:gtt}\\
-a^{0}(\theta')^{2}+\rho(\theta')^{2} & = & -1\label{gnn}
\end{eqnarray}
We also have third equation
\begin{eqnarray}
g^{t'n'} & = & -a(\theta')a^{0}(\theta')+\rho(\theta')\left(K(\theta')-\rho(\theta')\right)=0\label{eq:gtn}
\end{eqnarray}
Equations \eqref{eq:gtt} and \eqref{eq:gtn} give
\begin{eqnarray*}
a^{2}(a^{0})^{2}/\rho^{2}-a^{2} & = & 1\\
\implies-a^{0}(\theta')^{2}+\rho(\theta')^{2} & = & -\rho^{2}/a^{2}\\
\implies\rho^{2} & = & a^{2}\\
\implies a & = & \pm\rho
\end{eqnarray*}
In the third step we have used equation \eqref{gnn}. Substituting
this in equation \eqref{eq:gtt} gives
\begin{eqnarray*}
2\rho K & = & K^{2}-1\\
\implies\rho & = & \frac{1}{2}\left(K-\frac{1}{K}\right)
\end{eqnarray*}
Substituting this in equation \eqref{eq:gtn} gives
\begin{eqnarray*}
aa^{0} & = & \rho(K-\rho)\\
\implies a^{0} & = & \pm\frac{1}{2}\left(K+\frac{1}{K}\right)
\end{eqnarray*}
Collecting all the results
\begin{eqnarray*}
t & = & \pm\left(Kt'-\frac{1}{2}\left(K-\frac{1}{K}\right)(t'-n')\right)+c_{2}(\theta')\\
n & = & Kn'+\frac{1}{2}\left(K-\frac{1}{K}\right)(t'-n')+c_{1}(\theta')
\end{eqnarray*}
$\pm$ corresponds to just time reversal. So we choose only the $+$
solution. Undetermined functions $c_{1}(\theta'),c_{2}(\theta')$
are related to supertranslations. So the complete transformations
are
\begin{eqnarray}
t & = & \frac{t'}{2}\left(K+\frac{1}{K}\right)+\frac{n'}{2}\left(K-\frac{1}{K}\right)+c_{2}(\theta')+O(n'^{-p-1})\nonumber \\
n & = & \frac{n'}{2}\left(K+\frac{1}{K}\right)+\frac{t'}{2}\left(K-\frac{1}{K}\right)+c_{1}(\theta')+O(n'^{-p-1})\nonumber \\
\theta & = & 2\arctan\left(e^{-\nu}\tan\frac{\theta'}{2}\right)+O(n'^{-p-1});p\geq0\nonumber \\
\phi & = & \phi'\nonumber \\
K(\theta') & = & \cosh\nu+\cos\theta'\sinh\nu\label{eq:lorentz trans}
\end{eqnarray}
In the direction $\theta=\theta'=0$, above transformation is identical
to Lorentz transformation with boost $-\nu$. 

Some ideas in the above proof are similar to the derivation in the
Bondi-Burg-Metzner-Sachs (BMS) paper (part C of \citep{10.2307/2414436}).
However, there are important differences in motivations, assumptions,
flow of argument and result. BMS starts with Minkowski line element
with $O(1/r)$ corrections and discovers that it is invariant under
Lorentz transformation along with supertranslation. \textcolor{black}{Motivation
is to find out asymptotic symmetry group. Closely looking at the derivation
one finds that there are two independent parts. First is invariance
of Minkowski line element $ds^{2}=-du^{2}-2dud\tilde{r}$ under Lorentz
transformation with boost $\tilde{\nu}$. Second is invariance of
angular part of line element on sphere $ds^{2}=r^{2}d\theta^{2}+r^{2}\sin\theta^{2}d\phi^{2}$
under surface transformation parametrized by $\nu$. Identifying $\tilde{r}$
with $r$ implies $\tilde{\nu}=\nu$ . Gravity fixes some of the $1/r$
corrections which determines the mass transformation. }

\textcolor{black}{Our starting motivation is to find transformation
under which local information is invariant. So we demand invariance
of angular part of line element on sphere which is much weaker assumption.
Then we use the velocity from partition merger and determine mass
transformation. This is the keystone from which four-dimensional invariant
line element follows.}

\section{Observables\label{sec:Observables}}

\textcolor{black}{In this section we will explore the physical meaning
of the coefficients $r_{n^{2}}$ . This will help us find out partition
content of the asymptote. }

\subsection{\textcolor{black}{$r_{1}$\label{subsec:r_1}}}

\textcolor{black}{Let $L>>1$ denote the last element and $L_{k}$
denote the $k^{th}$ last element of the sequence $\left\{ 1,2,3,...,n\right\} $.
Similarly, let $f(L_{k})$ denote the $k^{th}$ last element of the
sequence $\left\{ f(M):M\in\left\{ m_{1},m_{2},...,m_{n}\right\} \right\} $.
For the analytic purpose we will treat $L\rightarrow\infty$. Consider
the limit $M\rightarrow L$.
\begin{eqnarray*}
\lim_{M\rightarrow L}P(M^{2}) & = & P_{1}(L^{2})
\end{eqnarray*}
So define
\begin{eqnarray}
r_{1} & \equiv & \lim_{M\rightarrow L}P(M^{2})e^{-G^{2}M^{2}}\label{eq:lim1}
\end{eqnarray}
which should match with the result obtained from calculating residue
\eqref{eq:lead residue}. Calculated numerically from residue, $r_{1}=1.001001011$
up to 9 decimal places. Explicit calculation of partition gives $P(n^{2}\geq16)=1.001001011\times10^{n^{2}}$
up to 9 decimal places. Using equation \eqref{eq:lim1} we get $r_{1}=1.001001011$ for $L=16$.
The two results match. Value of $L$ can be chosen based on desired
accuracy. Extrapolating to finite $m$ we get
\begin{eqnarray}
P_{1}(m^{2}) & = & \left(\lim_{M\rightarrow L}P(M^{2})e^{-G^{2}M^{2}}\right)e^{G^{2}m^{2}}\label{eq:extrapolate 1}
\end{eqnarray}
}This gives the amount of information contained in the leading asymptote.
Exact information depend on $L$. $r_{1}$ is numerically same as
$P(L^{2})e^{-G^{2}L^{2}}$. Except $e^{-G^{2}L^{2}}$ which is just
a multiplicative factor, information content is 
\begin{eqnarray*}
r_{1} & = & P\{L^{2}\}
\end{eqnarray*}
For example, if $L=16$ then information content of $r_{1}$ is partitions
of $P\{16^{2}\}$. 

\subsection{\textcolor{black}{$r_{2^{2}}$\label{subsec:r_4}}}

\textcolor{black}{For general $m^{2}$, equation \eqref{eq:extrapolate 1}
says that extrapolate $P_{1}(L^{2})$ backwards. This under-estimates
$P(m^{2})$ for $m<L$. In the limit $M\rightarrow(L-1)$ we get
\begin{eqnarray*}
\lim_{M\rightarrow L-1}\left(P(M^{2})-P_{1}(M^{2})\right) & = & P_{2}((L-1)^{2})
\end{eqnarray*}
Sequence $\left\{ \left(P(m^{2})-P_{1}(m^{2})\right)e^{-G^{2}m^{2}/4}\right\} $
has $t_{2}=2$ subsequences $s_{2}^{t}(m)\mid\{1,4,9,...\}$, $t\in T_{2}=\{1,2,...,t_{2}\}$
with different limits. 
\begin{eqnarray*}
\lim_{s_{2}^{t}(M)\rightarrow s_{2}^{t}(L)}\left(P(s_{2}^{t}(M))-P_{1}(s_{2}^{t}(M))\right) & = & P_{4}(s_{2}^{t}(L))
\end{eqnarray*}
For sufficiently large $L$, $P_{4}(s_{2}^{t}(L))$ matches with $P(s_{2}^{t}(L))-P_{1}(s_{2}^{t}(L))$
up-to desired accuracy . So define
\begin{eqnarray}
r_{4}^{t} & \equiv & \lim_{s_{2}^{t}(M)\rightarrow s_{2}^{t}(L)}\left(P(s_{2}^{t}(M))-P_{1}(s_{2}^{t}(M))\right)e^{-G^{2}s_{2}^{t}(M)/4}\label{eq:lim2}
\end{eqnarray}
Explicit calculation of partition for $n^{2}\geq64$ gives
\begin{eqnarray*}
P(n^{2}) & = & \begin{cases}
-0.011124359\times10^{m^{2}/4} & m=2k\\
-0.006260251\times10^{m^{2}/4} & m=2k+1
\end{cases}
\end{eqnarray*}
up to 9 decimal places. Using equation \eqref{eq:lim2} we get
\begin{eqnarray*}
r_{2^{2}}=r_{4} & = & \begin{cases}
-0.011124359 & m=2k\\
-0.006260251 & m=2k+1
\end{cases}
\end{eqnarray*}
This matches with the result obtained from calculating residue \eqref{eq:r4}.
Extrapolating to finite $m$
\begin{eqnarray}
P_{2}(m^{2}) & = & \left(\lim_{s_{2}^{t}(M)\rightarrow s_{2}^{t}(L)}\left(P(s_{2}^{t}(M))-P_{1}(s_{2}^{t}(M))\right)e^{-Gs_{2}^{t}(M)/4}\right)e^{G^{2}m^{2}/2^{2}}\label{eq:extrapolate 2}\\
t & = & [m^{2}]\in T_{2}\nonumber 
\end{eqnarray}
}where $[m^{2}]$ denote the subsequence in which $m^{2}$ appears.
This gives the amount of information contained in the second asymptote. 

\subsection{\textcolor{black}{$r_{s(1)}$}}

\textcolor{black}{Sequence $\{P(s(m))e^{-G^{2}s(m)/s(1)}\}$ may have
$t_{1}$ subsequences $s_{1}^{t}(m)\mid s(m)$, $t\in T_{1}=\{1,2,3,...,t_{1}\}$
with different limits.
\begin{eqnarray*}
\lim_{s(M)\rightarrow s_{1}^{t}(L)}P(s(M)) & = & P_{s(1)}(s_{1}^{t}(L)),t\in T_{1}
\end{eqnarray*}
So define
\begin{eqnarray}
r_{s(1)}^{t} & \equiv & \lim_{s_{1}^{t}(M)\rightarrow s_{1}^{t}(L)}P(s_{1}^{t}(M))e^{-G^{2}s_{1}^{t}(M)/s(1)},t\in T_{1}\label{eq:lim s(1)}
\end{eqnarray}
which should match with the result obtained from calculating residue.
One can see from the table \eqref{tab:Comparison-of-} that residue
matches with equation \eqref{eq:lim s(1)} for $m\geq3$. }
\begin{table}[tph]
\textcolor{black}{}%
\begin{tabular}{|c|c|c|}
\hline 
\textcolor{black}{$m$} & \textcolor{black}{$r_{s(1)}\times10^{s(m)/s(1)}$} & \textcolor{black}{$P(s(m))$}\tabularnewline
\hline 
\hline 
\textcolor{black}{1} & \textcolor{black}{$10.00000000100010000$} & \textcolor{black}{$10.00000000000000000$}\tabularnewline
\hline 
\textcolor{black}{2} & \textcolor{black}{$10.00000000100010001$} & \textcolor{black}{$10.00000000000000000$}\tabularnewline
\hline 
\textcolor{black}{3} & \textcolor{black}{$10.00000000100010000\times10^{15}$} & \textcolor{black}{$10.00000000100010000\times10^{15}$}\tabularnewline
\hline 
\textcolor{black}{4} & \textcolor{black}{$10.00000000100010001\times10^{19}$} & \textcolor{black}{$10.00000000100010001\times10^{19}$}\tabularnewline
\hline 
\textcolor{black}{5} & \textcolor{black}{$10.00000000100010000\times10^{43}$} & \textcolor{black}{$10.00000000100010000\times10^{43}$}\tabularnewline
\hline 
\textcolor{black}{6} & \textcolor{black}{$10.00000000100010001\times10^{65}$} & \textcolor{black}{$10.00000000100010001\times10^{65}$}\tabularnewline
\hline 
\textcolor{black}{7} & \textcolor{black}{$10.00000000100010001\times10^{79}$} & \textcolor{black}{$10.00000000100010001\times10^{79}$}\tabularnewline
\hline 
\textcolor{black}{8} & \textcolor{black}{$10.00000000100010000\times10^{115}$} & \textcolor{black}{$10.00000000100010000\times10^{115}$}\tabularnewline
\hline 
\textcolor{black}{9} & \textcolor{black}{$10.00000000100010001\times10^{157}$} & \textcolor{black}{$10.00000000100010001\times10^{157}$}\tabularnewline
\hline 
\end{tabular}\textcolor{black}{\caption{\label{tab:Comparison-of-}Comparison of $r_{s(1)}$ and partition
$P(s(1))$. Up to 17 decimal places they match for $m\protect\geq3$.}
}
\end{table}
\textcolor{black}{One can also see that up to 17 decimal places there
are two limit points. Extrapolating to finite $m$
\begin{eqnarray}
P_{s(1)}(s(m)) & = & \left(\lim_{s_{1}^{t}(M)\rightarrow s_{1}^{t}(L)}P(s_{1}^{t}(M))e^{-G^{2}s_{1}^{t}(M)/s(1)}\right)e^{G^{2}s(m)/s(1)}\label{eq:extrapolate s(1)}\\
t & = & [s(m)]\in T_{1}\nonumber 
\end{eqnarray}
}where $[s(m)]$ denote the subsequence in which $s(m)$ appears.

\subsection{\textcolor{black}{$r_{s(n+1)}$}}

\textcolor{black}{Assuming that $P_{s(n)}$ can be expressed as 
\begin{eqnarray*}
P_{s(n)}(s(m)) & = & \left(\lim_{s_{n}^{u}(M)\rightarrow s_{n}^{u}(L)}\left(P(s_{n}^{u}(M))-\sum_{j=1}^{n-1}P_{s(j)}(s_{n}^{u}(M))\right)e^{-G^{2}s_{n}^{u}(M)/s(n)}\right)e^{G^{2}s(m)/s(n)},\\
u & = & [s(m)]\in T_{n}=\left\{ 1,2,...,t_{n}\right\} 
\end{eqnarray*}
where $[s(m)]$ denotes the subsequence in which $s(m)$ appear. This
under-estimates $P(s_{n}^{u}(L_{k})),u\in T_{n}$ for $k>1$. 
\begin{eqnarray*}
\lim_{_{s_{n}^{u}(M)\rightarrow s_{n}^{u}(L_{2})}}\left(P(s_{n}^{u}(M))-\sum_{j=1}^{n}P_{s(j)}(s_{n}^{u}(M))\right) & = & P_{s(n+1)}\left(s_{n}^{u}(L_{2})\right),u\in T_{n}
\end{eqnarray*}
Sequence $\left\{ \left(P(s_{n}^{u}(M))-\sum_{j=1}^{n}P_{s(j)}(s_{n}^{u}(M))\right)e^{-G^{2}s_{n}^{u}(M)/s(n+1)}\right\} $
has $t_{n+1}$ subsequences $s_{n+1}^{t}(m)\mid s_{n}^{u}(m)$, $t\in T_{n+1}=\{1,2,...,t_{n+1}\}$
with different limits. 
\begin{eqnarray*}
\lim_{s_{n+1}^{t}(M)\rightarrow s_{n+1}^{t}(L)}\left(P(s_{n+1}^{t}(M))-\sum_{j=1}^{n}P_{s(j)}(s_{n+1}^{t}(m))\right) & = & P_{s(n+1)}(s_{n+1}^{t}(L)),t\in T_{n+1}
\end{eqnarray*}
This defines $r_{s(n+1)}^{t}$
\begin{eqnarray}
r_{s\left(n+1\right)}^{t} & \equiv & \lim_{s_{n+1}^{t}(M)\rightarrow s_{n+1}^{t}(L)}\left(P(s_{n+1}^{t}(M))-\sum_{j=1}^{n}P_{s(j)}(s_{n+1}^{t}(m))\right)e^{-G^{2}s_{n+1}^{t}(M)/s(n+1)}\label{eq:lim s(n)}
\end{eqnarray}
which should match with the result obtained from calculating residue.
Extrapolating to finite $m$
\begin{eqnarray}
P_{s(n+1)}(s(m)) & = & \left(\lim_{s_{n+1}^{t}(M)\rightarrow s_{n+1}^{t}(L)}\left(P(s_{n+1}^{t}(M))-\sum_{j=1}^{n}P_{s(j)}(s_{n+1}^{t}(M))\right)e^{-G^{2}s_{n+1}^{t}(M)/s(n+1)}\right)e^{G^{2}s(m)/s(n+1)}\nonumber \\
t & = & [s(m)]\in T_{n+1}=\{1,2,...,t_{n+1}\}\label{eq:extrapolate s(n)}
\end{eqnarray}
}

\subsection{Information and Radiation \label{sec:experiments}}

\textcolor{black}{$r_{s(n)}$ was defined in section \eqref{sec:Partition}
as
\begin{eqnarray}
P_{s(n)}(s(m)) & \equiv & -\sum_{i=1}^{s(n)}R(z_{i},s(m))\label{eq:P from finite}\\
r_{s(n)} & \equiv & P_{s(n)}(s(m))e^{-G^{2}s(m)/s(n)}\label{eq:r from finite}\\
R(z_{i},s(m)) & = & \lim_{z\rightarrow z_{0}}\frac{z-z_{i}}{z^{s(m)+1}}\prod_{l=1}^{\infty}\frac{1}{1-wz^{s(l)}}\nonumber \\
(1-wz_{i}^{s(n)}) & = & 0,\forall i\in\{1,2,...,s(n)\}\nonumber 
\end{eqnarray}
In general, $r_{s(n)}$ takes more than one value $\{r_{s(n)}^{1},r_{s(n)}^{2},...,r_{s(n)}^{t},...,r_{s(n)}^{t_{n}}\}$
where $t$ depends on $s(m)$.}\\
\textcolor{black}{Another way to define $r_{s(n)}^{t}$ is using equation
\eqref{eq:lim s(n)}. We motivated and also checked numerically the
definition \eqref{eq:extrapolate s(n)} matches with \eqref{eq:P from finite}
for finite $m$. While \eqref{eq:P from finite} and \eqref{eq:r from finite}
are calculated using finite number of initial terms of partition $\{P(s(1)),P(s(2)),....,P(s(m))\}$.
\eqref{eq:lim s(n)} and \eqref{eq:extrapolate s(n)} are calculated
using finite number of end terms of partition $\{P(s(L)),P(s(L_{2})),....,P(s(L_{m}))\}$.
These definitions extrapolated to finite $m$ match with each other.}

\textcolor{black}{The second definition has the advantage of being
physically meaningful. It tells us that $r_{n^{2}}$ is nothing but
partition. Along with this knowledge, $\Delta P_{n}$ in equation
\eqref{eq:info change} gives observable information. Since this observable
information is a result of change in background metric, it is nothing
but bremsstrahlung radiation. Still some work is necessary to find
out the exact form of the gravitational radiation. That will be done
in a follow up paper.}

\section{Conclusion\label{sec:Conclusion}}

\textcolor{black}{We have considered a model based on partition of
squares $\{1,4,9,...,m^{2},...:m\in\mathbb{{N}}\}$ with weight $w=e^{G^{2}}$
as defined in section \eqref{sec:Partition}. It can be expressed
as sum of asymptotes 
\begin{eqnarray*}
P\left(m^{2}\right) & = & \sum_{n\geq1}P_{n}(m^{2})\\
P_{n}(m^{2}) & = & r_{n^{2}}e^{G^{2}m^{2}/n^{2}}\\
r_{w,n^{2}} & = & \frac{1}{n^{2}}\prod_{k\neq n}\sum_{j=1}^{n^{2}}\left(\frac{e^{2\pi j\frac{m^{2}}{n^{2}}i}}{1-e^{2\pi j\frac{k^{2}}{n^{2}}i}w^{1-\frac{k^{2}}{n^{2}}}}\right)
\end{eqnarray*}
Leading asymptote grows like $e^{G^{2}m^{2}}$ and $n^{th}$ subleading
asymptote grows like $e^{G^{2}m^{2}/n^{2}}$ . There are two different
regimes. First is small $n$ regime $n<Gm$. This regime is closer
to partition perspective and most visible in the graph \eqref{fig:sq asymptotes}.
The coefficient $r_{1}$ has simple physical interpretation in terms
of partition as described in section \eqref{sec:Observables}. On
the other hand since $e^{G^{2}m^{2}/n^{2}}>1$, perturbative analysis
is not possible. Second is large $n$ regime $n>>Gm$. This regime
represent fine deviations of partition from the leading asymptotic
behavior and is difficult to see in the graph. Coefficients $r_{n^{2}}$
become more and more complicated for $n>>Gm$. On the other hand,
$e^{G^{2}m^{2}/n^{2}}\sim1\implies\frac{\Delta P_{n}}{\Delta(Gm)}=r_{n^{2}}\left(\frac{2G^{2}m}{n^{2}}\right)e^{G^{2}m^{2}/n^{2}}<<1$.
So perturbative analysis is possible. }

\textcolor{black}{We have studied merger of two partitions and realized
that the amount of merger can be used as a measure to define }\textcolor{black}{\emph{distance}}\textcolor{black}{{}
$n$. 
\begin{eqnarray*}
P\{m^{2}\}\bullet P\{M^{2}\} & \rightarrow P\{m^{2},M^{2},n\}\rightarrow & P\{(M+m)^{2}\}
\end{eqnarray*}
such that $P\{m^{2},M^{2},n=\infty\}=P\{m^{2}\}\bullet P\{M^{2}\}$
(two systems are completely separate) and $P\{m^{2},M^{2},n=0\}=P\{(M+m)^{2}\}$
(systems have completely merged). When intermediate state $P\{m^{2},M^{2},n\}$
is identified with the $n^{th}$ subleading asymptote then merger
matches with particle falling in black hole. Space and time acquires
meaning only in the context of merger. Prior to merger parts are observable
which are called }\textit{\textcolor{black}{space}}\textcolor{black}{.
Upon merger, space is converted to non-observable parts and information
is released. As a result, space shrinks with velocity $v=\sqrt{\frac{2GM}{n}}$
at radius $n$ from the black hole and test particle drags along with
the space fabric. }

\textcolor{black}{Effective physical picture emerges where a black
hole of mass $M$ acts as space sink and annihilates space of volume
$2GM$ per unit time. Specially $\frac{1}{n^{2}}$ factor in $\Delta v=\frac{2GM}{n^{2}}\Delta t$
would give an impression that there is conservation and continuity
relation. Space is flowing towards black hole and finally draining
into it. However information emitted during merger from surface of
radius $n$
\begin{eqnarray*}
\Delta P_{n} & = & r_{n^{2}}\left(\frac{2G^{2}Mm}{n^{2}}\right)e^{G^{2}M^{2}/n^{2}}
\end{eqnarray*}
reveals that the effective picture is misleading. Discussion in section
\eqref{sec:Observables} shows that coefficients $r_{n^{2}}$ are
nothing but partitions in convoluted form. Thus there is no analytic
relation between $r_{n^{2}}$ and $r_{(n+1)^{2}}$. To get $\Delta P_{n}$
one has to fall back to the definition of partition. In other words,
information on surface of radius $n$ and $(n+1)$ is not related.
So there is no flow of space from one surface to the next. Appearance
of flow is only emergent. Information released at two consecutive
time slices are unrelated. This is how causality emerges. }Emergent
causality is the guiding principle throughout the work. Place where
emergent patterns are found is Natural numbers or some derivative
of it. So we call this framework \textit{Sankhyaa} (pronounced Sunkh-ya).

\textcolor{black}{Majority of the information is contained in small
$n$ asymptotes. Information appears to be localized in a bounded
region which effectively looks like an event horizon. }Region inside
and far outside the horizon is described by small $n$ and large $n$
asymptotes respectively. \textcolor{black}{The sharp notion of event
horizon disappears and information is spread all over the space. This
unifies black hole and background spacetime and replaces them with
one fundamental object described by partition. We call it blackhole-space.
Small $n$ regime is closer to black-hole entropy and large $n$ regime
is closer to spacetime perspective. Important take away is that one
mechanism explains black-hole entropy and metric. This shows that
gravitational radiation carries black-hole entropy worth of information.
By analyzing the radiation we can retrieve the information emitted
during merger. In a follow up paper we will give corrections to Schwarzschild
metric and gravitational radiation from which information can be retrieved
upto $O(G^{3})$. Second prediction is that if radiation with right
information is sent in, instead of increasing mass of black hole it
will de-merge the black hole. This is reverse process of black-hole
formation. That is how, a free particle is able to resist fall by
boosting in opposite direction. We call this phenomenon as information
pressure. This will be shown in future work. }

From the above model it is clear that classical gravity cannot be
separated from quantum gravity. Gravity is quantized from the get-go.
The root lies in common origin of metric (large $n$ regime) and black-hole
entropy (small $n$ regime). Metric is usually assumed to have classical
meaning as in classical general relativity, while entropy is purely
quantum object without any classical description. This work shows
that metric and entropy are on equal footing. There is no meaning
of metric without entropy. This was expected a-priori for at least
couple of reasons. First reason comes from information puzzle. Black-hole
entropy is of quantum origin. On the other hand soft radiation is
known to carry to some information \citep{Chatterjee:2017zeb,Hawking:2016msc}.
This is classical in nature. Hence there is some ambiguity between
classical and quantum origin of information. Secondly, in emergent
causality, boundary condition is built into the system. This can also
be seen as unification of boundary condition with the differential
equation. Any attempt to unify differential equation with boundary
conditions will quantize the theory. Hence the system is quantized
by construction. This is one of the hallmarks of quantum gravity. 

\label{holography}A comprehensive discussion of information paradox
and associated confusions and subtleties can be found in \citep{Harlow:2014yka,Mathur:2009hf}.
While the spectrum of approaches is very broad, all of them have one
feature in common. They all interpret information as quantum information.
That is, all the approaches are based on qubits / quantum fields /
wavefunction which we will collectively refer to as field. Original
Hawking\textquoteright s argument of mixed nature of radiation is
also formulated in terms of quantum field. Field by definition requires
a pre-notion of spacetime. In some sense, the knot of information
paradox lies at lack of correct understanding of what do we mean by
information? 

Information content and interpretation is very different for a bit
(classical information) and a qubit. While separate classical bits
describe position and momentum, quantum field contains information
about both position and momentum. Bit cannot be used to describe qubit.
Just like qubit and bit are two completely different ways of describing
nature. In the same way, partition of natural numbers is a completely
new way of describing nature. Separate quantum fields are necessary
to describe matter and metric. Partition integrates information about
the particle and the spacetime by unifying them. Bits or qubits cannot
be used to describe partition. Instead of wavefunction of black hole,
partition of blackhole-space is more apt description of nature. This
work can also be seen as an interpretation of partition. We found
the prescription through which classical information emerges from
partition. More detailed study is necessary to see how field emerges
from partition. 

Seen as scattering process, Hawking radiation is a spontaneous decay
process of black hole. Above model allows de-merger of black hole
if right information is sent into the black hole. This suggests a
mechanism by which energy can flow out of black hole. In future work
we would like to explore these issues. 

\label{stringbh}There is also a volume of work on stringy microstates
of black hole summed up in \citep{Sen:2014aja}. These microstates count the number
of excitations of string in weak coupling limit. In perturbative string theory, strings or
branes are D-dimensional fundamental objects leaving in spacetime. Partition or
blackhole-space is also an extended fundamental object by itself. But it does not exist
inside spacetime. It is unified description of spacetime and matter
together. It is different from counting string
states. Partition is like degrees of freedom in natural numbers. At
best one can think of the parts as excitation of gravitons or degrees of freedom
of spacetime. Probably this work will have natural connection with non-perturbative
description of string theory.

\textcolor{black}{Experiments in physics are about observing field
and radiation far from the particle. This is same as measuring large
$n$ asymptotes of the partition. From that we reconstruct properties
of the particle. This is the only known way of observing nature till
now. We do not see the particle as whole. }This work suggests that
there could be experiments for small $n$ regime as well, which would
be new class of experiments closer to partition perspective and far
from spacetime perspective. A stronger claim is that it could be possible
to see the whole particle if one can capture the full partition $P\{n^{2}\}$
(this is exact partition and not just the leading asymptotes) in one
go.\textcolor{black}{{} This would be completely new way of observing
nature devoid of space and time. Experimental aspects of partition
will be demonstrated in future project. }
\begin{acknowledgments}
I would like to thank Ashoke Sen for number of discussions. I also
acknowledge the support of Brown University, USA and hospitality of
Vedanta Society of Providence, USA where initial part of the work
was done. Most of the work is motivated by numerous insightful discussions
with Swami Yogatmananda at the Vedanta Society. 
\end{acknowledgments}

\bibliographystyle{apsrev4-1}
\bibliography{reference}

\begin{thebibliography}{37}%
\makeatletter
\providecommand \@ifxundefined [1]{%
 \@ifx{#1\undefined}
}%
\providecommand \@ifnum [1]{%
 \ifnum #1\expandafter \@firstoftwo
 \else \expandafter \@secondoftwo
 \fi
}%
\providecommand \@ifx [1]{%
 \ifx #1\expandafter \@firstoftwo
 \else \expandafter \@secondoftwo
 \fi
}%
\providecommand \natexlab [1]{#1}%
\providecommand \enquote  [1]{``#1''}%
\providecommand \bibnamefont  [1]{#1}%
\providecommand \bibfnamefont [1]{#1}%
\providecommand \citenamefont [1]{#1}%
\providecommand \href@noop [0]{\@secondoftwo}%
\providecommand \href [0]{\begingroup \@sanitize@url \@href}%
\providecommand \@href[1]{\@@startlink{#1}\@@href}%
\providecommand \@@href[1]{\endgroup#1\@@endlink}%
\providecommand \@sanitize@url [0]{\catcode `\\12\catcode `\$12\catcode
  `\&12\catcode `\#12\catcode `\^12\catcode `\_12\catcode `\%12\relax}%
\providecommand \@@startlink[1]{}%
\providecommand \@@endlink[0]{}%
\providecommand \url  [0]{\begingroup\@sanitize@url \@url }%
\providecommand \@url [1]{\endgroup\@href {#1}{\urlprefix }}%
\providecommand \urlprefix  [0]{URL }%
\providecommand \Eprint [0]{\href }%
\providecommand \doibase [0]{http://dx.doi.org/}%
\providecommand \selectlanguage [0]{\@gobble}%
\providecommand \bibinfo  [0]{\@secondoftwo}%
\providecommand \bibfield  [0]{\@secondoftwo}%
\providecommand \translation [1]{[#1]}%
\providecommand \BibitemOpen [0]{}%
\providecommand \bibitemStop [0]{}%
\providecommand \bibitemNoStop [0]{.\EOS\space}%
\providecommand \EOS [0]{\spacefactor3000\relax}%
\providecommand \BibitemShut  [1]{\csname bibitem#1\endcsname}%
\let\auto@bib@innerbib\@empty
\bibitem [{\citenamefont {Jannes}(2009)}]{Jannes:2009yr}%
  \BibitemOpen
  \bibfield  {author} {\bibinfo {author} {\bibfnamefont {G.}~\bibnamefont
  {Jannes}},\ }\emph {\bibinfo {title} {{Emergent gravity: the BEC
  paradigm}}},\ \href@noop {} {Ph.D. thesis},\ \bibinfo  {school} {Madrid U.}
  (\bibinfo {year} {2009}),\ \Eprint {http://arxiv.org/abs/0907.2839}
  {arXiv:0907.2839 [gr-qc]} \BibitemShut {NoStop}%
\bibitem [{\citenamefont {Finazzi}\ \emph {et~al.}(2012)\citenamefont
  {Finazzi}, \citenamefont {Liberati},\ and\ \citenamefont
  {Sindoni}}]{Finazzi:2011zw}%
  \BibitemOpen
  \bibfield  {author} {\bibinfo {author} {\bibfnamefont {S.}~\bibnamefont
  {Finazzi}}, \bibinfo {author} {\bibfnamefont {S.}~\bibnamefont {Liberati}}, \
  and\ \bibinfo {author} {\bibfnamefont {L.}~\bibnamefont {Sindoni}},\ }\href
  {\doibase 10.1103/PhysRevLett.108.071101} {\bibfield  {journal} {\bibinfo
  {journal} {Phys. Rev. Lett.}\ }\textbf {\bibinfo {volume} {108}},\ \bibinfo
  {pages} {071101} (\bibinfo {year} {2012})},\ \Eprint
  {http://arxiv.org/abs/1103.4841} {arXiv:1103.4841 [gr-qc]} \BibitemShut
  {NoStop}%
\bibitem [{\citenamefont {Vancea}\ and\ \citenamefont
  {Santos}(2012)}]{Vancea:2010vf}%
  \BibitemOpen
  \bibfield  {author} {\bibinfo {author} {\bibfnamefont {I.~V.}\ \bibnamefont
  {Vancea}}\ and\ \bibinfo {author} {\bibfnamefont {M.~A.}\ \bibnamefont
  {Santos}},\ }\href {\doibase 10.1142/S0217732312500125} {\bibfield  {journal}
  {\bibinfo  {journal} {Mod. Phys. Lett.}\ }\textbf {\bibinfo {volume} {A27}},\
  \bibinfo {pages} {1250012} (\bibinfo {year} {2012})},\ \Eprint
  {http://arxiv.org/abs/1002.2454} {arXiv:1002.2454 [hep-th]} \BibitemShut
  {NoStop}%
\bibitem [{\citenamefont {Verlinde}(2011)}]{Verlinde:2010hp}%
  \BibitemOpen
  \bibfield  {author} {\bibinfo {author} {\bibfnamefont {E.~P.}\ \bibnamefont
  {Verlinde}},\ }\href {\doibase 10.1007/JHEP04(2011)029} {\bibfield  {journal}
  {\bibinfo  {journal} {JHEP}\ }\textbf {\bibinfo {volume} {04}},\ \bibinfo
  {pages} {029} (\bibinfo {year} {2011})},\ \Eprint
  {http://arxiv.org/abs/1001.0785} {arXiv:1001.0785 [hep-th]} \BibitemShut
  {NoStop}%
\bibitem [{\citenamefont {Verlinde}(2017)}]{Verlinde:2016toy}%
  \BibitemOpen
  \bibfield  {author} {\bibinfo {author} {\bibfnamefont {E.~P.}\ \bibnamefont
  {Verlinde}},\ }\href {\doibase 10.21468/SciPostPhys.2.3.016} {\bibfield
  {journal} {\bibinfo  {journal} {SciPost Phys.}\ }\textbf {\bibinfo {volume}
  {2}},\ \bibinfo {pages} {016} (\bibinfo {year} {2017})},\ \Eprint
  {http://arxiv.org/abs/1611.02269} {arXiv:1611.02269 [hep-th]} \BibitemShut
  {NoStop}%
\bibitem [{\citenamefont {Yang}(2008)}]{Yang:2007qx}%
  \BibitemOpen
  \bibfield  {author} {\bibinfo {author} {\bibfnamefont {H.~S.}\ \bibnamefont
  {Yang}},\ }\bibfield  {booktitle} {\emph {\bibinfo {booktitle} {{Lie theory
  and its applications in physics. Proceedings, 7th International Workshop,
  Varna, Bulgaria, June 18-24, 2007}}},\ }\href@noop {} {\bibfield  {journal}
  {\bibinfo  {journal} {Bulg. J. Phys.}\ }\textbf {\bibinfo {volume} {35}},\
  \bibinfo {pages} {323} (\bibinfo {year} {2008})},\ \Eprint
  {http://arxiv.org/abs/0711.0234} {arXiv:0711.0234 [hep-th]} \BibitemShut
  {NoStop}%
\bibitem [{\citenamefont {Rivelles}(2011)}]{Rivelles:2011gq}%
  \BibitemOpen
  \bibfield  {author} {\bibinfo {author} {\bibfnamefont {V.~O.}\ \bibnamefont
  {Rivelles}},\ }\bibfield  {booktitle} {\emph {\bibinfo {booktitle}
  {{Proceedings, 14th Mexican School of Particles and Fields (MSPF 2010):
  Morelia, Mexico, November 4-12, 2010}}},\ }\href {\doibase
  10.1088/1742-6596/287/1/012012} {\bibfield  {journal} {\bibinfo  {journal}
  {J. Phys. Conf. Ser.}\ }\textbf {\bibinfo {volume} {287}},\ \bibinfo {pages}
  {012012} (\bibinfo {year} {2011})},\ \Eprint {http://arxiv.org/abs/1101.4579}
  {arXiv:1101.4579 [hep-th]} \BibitemShut {NoStop}%
\bibitem [{\citenamefont {Lloyd}(2005)}]{Lloyd:2005js}%
  \BibitemOpen
  \bibfield  {author} {\bibinfo {author} {\bibfnamefont {S.}~\bibnamefont
  {Lloyd}},\ }\href@noop {} {\bibfield  {journal} {\bibinfo  {journal}
  {Submitted to: Science}\ } (\bibinfo {year} {2005})},\ \Eprint
  {http://arxiv.org/abs/quant-ph/0501135} {arXiv:quant-ph/0501135 [quant-ph]}
  \BibitemShut {NoStop}%
\bibitem [{\citenamefont {Steinacker}(2010)}]{Steinacker:2010rh}%
  \BibitemOpen
  \bibfield  {author} {\bibinfo {author} {\bibfnamefont {H.}~\bibnamefont
  {Steinacker}},\ }\href {\doibase 10.1088/0264-9381/27/13/133001} {\bibfield
  {journal} {\bibinfo  {journal} {Class. Quant. Grav.}\ }\textbf {\bibinfo
  {volume} {27}},\ \bibinfo {pages} {133001} (\bibinfo {year} {2010})},\
  \Eprint {http://arxiv.org/abs/1003.4134} {arXiv:1003.4134 [hep-th]}
  \BibitemShut {NoStop}%
\bibitem [{\citenamefont {Almheiri}\ \emph {et~al.}(2015)\citenamefont
  {Almheiri}, \citenamefont {Dong},\ and\ \citenamefont
  {Harlow}}]{Almheiri:2014lwa}%
  \BibitemOpen
  \bibfield  {author} {\bibinfo {author} {\bibfnamefont {A.}~\bibnamefont
  {Almheiri}}, \bibinfo {author} {\bibfnamefont {X.}~\bibnamefont {Dong}}, \
  and\ \bibinfo {author} {\bibfnamefont {D.}~\bibnamefont {Harlow}},\ }\href
  {\doibase 10.1007/JHEP04(2015)163} {\bibfield  {journal} {\bibinfo  {journal}
  {JHEP}\ }\textbf {\bibinfo {volume} {04}},\ \bibinfo {pages} {163} (\bibinfo
  {year} {2015})},\ \Eprint {http://arxiv.org/abs/1411.7041} {arXiv:1411.7041
  [hep-th]} \BibitemShut {NoStop}%
\bibitem [{\citenamefont {Harlow}(2017)}]{Harlow:2016vwg}%
  \BibitemOpen
  \bibfield  {author} {\bibinfo {author} {\bibfnamefont {D.}~\bibnamefont
  {Harlow}},\ }\href {\doibase 10.1007/s00220-017-2904-z} {\bibfield  {journal}
  {\bibinfo  {journal} {Commun. Math. Phys.}\ }\textbf {\bibinfo {volume}
  {354}},\ \bibinfo {pages} {865} (\bibinfo {year} {2017})},\ \Eprint
  {http://arxiv.org/abs/1607.03901} {arXiv:1607.03901 [hep-th]} \BibitemShut
  {NoStop}%
\bibitem [{\citenamefont {Horowitz}\ and\ \citenamefont
  {Polchinski}(2006)}]{Horowitz:2006ct}%
  \BibitemOpen
  \bibfield  {author} {\bibinfo {author} {\bibfnamefont {G.~T.}\ \bibnamefont
  {Horowitz}}\ and\ \bibinfo {author} {\bibfnamefont {J.}~\bibnamefont
  {Polchinski}},\ }\href@noop {} {\ ,\ \bibinfo {pages} {169} (\bibinfo {year}
  {2006})},\ \Eprint {http://arxiv.org/abs/gr-qc/0602037} {arXiv:gr-qc/0602037
  [gr-qc]} \BibitemShut {NoStop}%
\bibitem [{\citenamefont {Kovtun}\ \emph {et~al.}(2005)\citenamefont {Kovtun},
  \citenamefont {Son},\ and\ \citenamefont {Starinets}}]{Kovtun:2004de}%
  \BibitemOpen
  \bibfield  {author} {\bibinfo {author} {\bibfnamefont {P.}~\bibnamefont
  {Kovtun}}, \bibinfo {author} {\bibfnamefont {D.~T.}\ \bibnamefont {Son}}, \
  and\ \bibinfo {author} {\bibfnamefont {A.~O.}\ \bibnamefont {Starinets}},\
  }\href {\doibase 10.1103/PhysRevLett.94.111601} {\bibfield  {journal}
  {\bibinfo  {journal} {Phys. Rev. Lett.}\ }\textbf {\bibinfo {volume} {94}},\
  \bibinfo {pages} {111601} (\bibinfo {year} {2005})},\ \Eprint
  {http://arxiv.org/abs/hep-th/0405231} {arXiv:hep-th/0405231 [hep-th]}
  \BibitemShut {NoStop}%
\bibitem [{\citenamefont {Heckman}\ and\ \citenamefont
  {Verlinde}(2011)}]{Heckman:2011qu}%
  \BibitemOpen
  \bibfield  {author} {\bibinfo {author} {\bibfnamefont {J.~J.}\ \bibnamefont
  {Heckman}}\ and\ \bibinfo {author} {\bibfnamefont {H.}~\bibnamefont
  {Verlinde}},\ }\href@noop {} {\  (\bibinfo {year} {2011})},\ \Eprint
  {http://arxiv.org/abs/1112.5210} {arXiv:1112.5210 [hep-th]} \BibitemShut
  {NoStop}%
\bibitem [{\citenamefont {Carlip}(2014)}]{Carlip:2012wa}%
  \BibitemOpen
  \bibfield  {author} {\bibinfo {author} {\bibfnamefont {S.}~\bibnamefont
  {Carlip}},\ }\href {\doibase 10.1016/j.shpsb.2012.11.002} {\bibfield
  {journal} {\bibinfo  {journal} {Stud. Hist. Phil. Sci.}\ }\textbf {\bibinfo
  {volume} {B46}},\ \bibinfo {pages} {200} (\bibinfo {year} {2014})},\ \Eprint
  {http://arxiv.org/abs/1207.2504} {arXiv:1207.2504 [gr-qc]} \BibitemShut
  {NoStop}%
\bibitem [{\citenamefont {Marolf}(2015)}]{Marolf:2014yga}%
  \BibitemOpen
  \bibfield  {author} {\bibinfo {author} {\bibfnamefont {D.}~\bibnamefont
  {Marolf}},\ }\href {\doibase 10.1103/PhysRevLett.114.031104} {\bibfield
  {journal} {\bibinfo  {journal} {Phys. Rev. Lett.}\ }\textbf {\bibinfo
  {volume} {114}},\ \bibinfo {pages} {031104} (\bibinfo {year} {2015})},\
  \Eprint {http://arxiv.org/abs/1409.2509} {arXiv:1409.2509 [hep-th]}
  \BibitemShut {NoStop}%
\bibitem [{\citenamefont {Giddings}\ and\ \citenamefont
  {Strominger}(1988)}]{Giddings:1988cx}%
  \BibitemOpen
  \bibfield  {author} {\bibinfo {author} {\bibfnamefont {S.~B.}\ \bibnamefont
  {Giddings}}\ and\ \bibinfo {author} {\bibfnamefont {A.}~\bibnamefont
  {Strominger}},\ }\href {\doibase 10.1016/0550-3213(88)90109-5} {\bibfield
  {journal} {\bibinfo  {journal} {Nucl. Phys.}\ }\textbf {\bibinfo {volume}
  {B307}},\ \bibinfo {pages} {854} (\bibinfo {year} {1988})}\BibitemShut
  {NoStop}%
\bibitem [{\citenamefont {Hartle}\ and\ \citenamefont
  {Hawking}(1983)}]{Hartle:1983ai}%
  \BibitemOpen
  \bibfield  {author} {\bibinfo {author} {\bibfnamefont {J.~B.}\ \bibnamefont
  {Hartle}}\ and\ \bibinfo {author} {\bibfnamefont {S.~W.}\ \bibnamefont
  {Hawking}},\ }\href {\doibase 10.1103/PhysRevD.28.2960} {\bibfield  {journal}
  {\bibinfo  {journal} {Phys. Rev.}\ }\textbf {\bibinfo {volume} {D28}},\
  \bibinfo {pages} {2960} (\bibinfo {year} {1983})},\ \bibinfo {note} {[Adv.
  Ser. Astrophys. Cosmol.3,174(1987)]}\BibitemShut {NoStop}%
\bibitem [{\citenamefont {Hartle}(1986)}]{Hartle:1986eu}%
  \BibitemOpen
  \bibfield  {author} {\bibinfo {author} {\bibfnamefont {J.~B.}\ \bibnamefont
  {Hartle}},\ }in\ \href@noop {} {\emph {\bibinfo {booktitle} {{27th Liege
  International Astrophysical Colloquium on Origin and Ea Early History of the
  Universe Liege, Belgium, July 1-4, 1986}}}}\ (\bibinfo {year} {1986})\ pp.\
  \bibinfo {pages} {1--19},\ \bibinfo {note} {[,1(1986)]}\BibitemShut {NoStop}%
\bibitem [{\citenamefont {Wudka}(1987)}]{Wudka:1987ef}%
  \BibitemOpen
  \bibfield  {author} {\bibinfo {author} {\bibfnamefont {J.}~\bibnamefont
  {Wudka}},\ }\href {\doibase 10.1103/PhysRevD.36.1036} {\bibfield  {journal}
  {\bibinfo  {journal} {Phys. Rev.}\ }\textbf {\bibinfo {volume} {D36}},\
  \bibinfo {pages} {1036} (\bibinfo {year} {1987})}\BibitemShut {NoStop}%
\bibitem [{\citenamefont {Harlow}(2016)}]{Harlow:2014yka}%
  \BibitemOpen
  \bibfield  {author} {\bibinfo {author} {\bibfnamefont {D.}~\bibnamefont
  {Harlow}},\ }\href {\doibase 10.1103/RevModPhys.88.015002} {\bibfield
  {journal} {\bibinfo  {journal} {Rev. Mod. Phys.}\ }\textbf {\bibinfo {volume}
  {88}},\ \bibinfo {pages} {015002} (\bibinfo {year} {2016})},\ \Eprint
  {http://arxiv.org/abs/1409.1231} {arXiv:1409.1231 [hep-th]} \BibitemShut
  {NoStop}%
\bibitem [{\citenamefont {Giddings}(1995)}]{Giddings:1995gd}%
  \BibitemOpen
  \bibfield  {author} {\bibinfo {author} {\bibfnamefont {S.~B.}\ \bibnamefont
  {Giddings}},\ }in\ \href@noop {} {\emph {\bibinfo {booktitle} {{Particles,
  strings and cosmology. Proceedings, 19th Johns Hopkins Workshop and 5th
  PASCOS Interdisciplinary Symposium, Baltimore, USA, March 22-25, 1995}}}}\
  (\bibinfo {year} {1995})\ pp.\ \bibinfo {pages} {415--428},\ \Eprint
  {http://arxiv.org/abs/hep-th/9508151} {arXiv:hep-th/9508151 [hep-th]}
  \BibitemShut {NoStop}%
\bibitem [{\citenamefont {Mathur}(2009)}]{Mathur:2009hf}%
  \BibitemOpen
  \bibfield  {author} {\bibinfo {author} {\bibfnamefont {S.~D.}\ \bibnamefont
  {Mathur}},\ }\bibfield  {booktitle} {\emph {\bibinfo {booktitle} {{Strings,
  Supergravity and Gauge Theories. Proceedings, CERN Winter School, CERN,
  Geneva, Switzerland, February 9-13 2009}}},\ }\href {\doibase
  10.1088/0264-9381/26/22/224001} {\bibfield  {journal} {\bibinfo  {journal}
  {Class. Quant. Grav.}\ }\textbf {\bibinfo {volume} {26}},\ \bibinfo {pages}
  {224001} (\bibinfo {year} {2009})},\ \Eprint {http://arxiv.org/abs/0909.1038}
  {arXiv:0909.1038 [hep-th]} \BibitemShut {NoStop}%
\bibitem [{\citenamefont {Hawking}\ \emph {et~al.}(2016)\citenamefont
  {Hawking}, \citenamefont {Perry},\ and\ \citenamefont
  {Strominger}}]{Hawking:2016msc}%
  \BibitemOpen
  \bibfield  {author} {\bibinfo {author} {\bibfnamefont {S.~W.}\ \bibnamefont
  {Hawking}}, \bibinfo {author} {\bibfnamefont {M.~J.}\ \bibnamefont {Perry}},
  \ and\ \bibinfo {author} {\bibfnamefont {A.}~\bibnamefont {Strominger}},\
  }\href {\doibase 10.1103/PhysRevLett.116.231301} {\bibfield  {journal}
  {\bibinfo  {journal} {Phys. Rev. Lett.}\ }\textbf {\bibinfo {volume} {116}},\
  \bibinfo {pages} {231301} (\bibinfo {year} {2016})},\ \Eprint
  {http://arxiv.org/abs/1601.00921} {arXiv:1601.00921 [hep-th]} \BibitemShut
  {NoStop}%
\bibitem [{\citenamefont {Haco}\ \emph {et~al.}(2018)\citenamefont {Haco},
  \citenamefont {Hawking}, \citenamefont {Perry},\ and\ \citenamefont
  {Strominger}}]{Haco:2018ske}%
  \BibitemOpen
  \bibfield  {author} {\bibinfo {author} {\bibfnamefont {S.}~\bibnamefont
  {Haco}}, \bibinfo {author} {\bibfnamefont {S.~W.}\ \bibnamefont {Hawking}},
  \bibinfo {author} {\bibfnamefont {M.~J.}\ \bibnamefont {Perry}}, \ and\
  \bibinfo {author} {\bibfnamefont {A.}~\bibnamefont {Strominger}},\ }\href
  {\doibase 10.1007/JHEP12(2018)098} {\bibfield  {journal} {\bibinfo  {journal}
  {JHEP}\ }\textbf {\bibinfo {volume} {12}},\ \bibinfo {pages} {098} (\bibinfo
  {year} {2018})},\ \Eprint {http://arxiv.org/abs/1810.01847} {arXiv:1810.01847
  [hep-th]} \BibitemShut {NoStop}%
\bibitem [{\citenamefont {Susskind}(2016)}]{Susskind:2014rva}%
  \BibitemOpen
  \bibfield  {author} {\bibinfo {author} {\bibfnamefont {L.}~\bibnamefont
  {Susskind}},\ }\href {\doibase 10.1002/prop.201500093,
  10.1002/prop.201500092} {\bibfield  {journal} {\bibinfo  {journal} {Fortsch.
  Phys.}\ }\textbf {\bibinfo {volume} {64}},\ \bibinfo {pages} {44} (\bibinfo
  {year} {2016})},\ \bibinfo {note} {[Fortsch. Phys.64,24(2016)]},\ \Eprint
  {http://arxiv.org/abs/1403.5695} {arXiv:1403.5695 [hep-th]} \BibitemShut
  {NoStop}%
\bibitem [{\citenamefont {Faulkner}\ \emph {et~al.}(2014)\citenamefont
  {Faulkner}, \citenamefont {Guica}, \citenamefont {Hartman}, \citenamefont
  {Myers},\ and\ \citenamefont {Van~Raamsdonk}}]{Faulkner:2013ica}%
  \BibitemOpen
  \bibfield  {author} {\bibinfo {author} {\bibfnamefont {T.}~\bibnamefont
  {Faulkner}}, \bibinfo {author} {\bibfnamefont {M.}~\bibnamefont {Guica}},
  \bibinfo {author} {\bibfnamefont {T.}~\bibnamefont {Hartman}}, \bibinfo
  {author} {\bibfnamefont {R.~C.}\ \bibnamefont {Myers}}, \ and\ \bibinfo
  {author} {\bibfnamefont {M.}~\bibnamefont {Van~Raamsdonk}},\ }\href {\doibase
  10.1007/JHEP03(2014)051} {\bibfield  {journal} {\bibinfo  {journal} {JHEP}\
  }\textbf {\bibinfo {volume} {03}},\ \bibinfo {pages} {051} (\bibinfo {year}
  {2014})},\ \Eprint {http://arxiv.org/abs/1312.7856} {arXiv:1312.7856
  [hep-th]} \BibitemShut {NoStop}%
\bibitem [{\citenamefont {Nishioka}\ \emph {et~al.}(2009)\citenamefont
  {Nishioka}, \citenamefont {Ryu},\ and\ \citenamefont
  {Takayanagi}}]{Nishioka:2009un}%
  \BibitemOpen
  \bibfield  {author} {\bibinfo {author} {\bibfnamefont {T.}~\bibnamefont
  {Nishioka}}, \bibinfo {author} {\bibfnamefont {S.}~\bibnamefont {Ryu}}, \
  and\ \bibinfo {author} {\bibfnamefont {T.}~\bibnamefont {Takayanagi}},\
  }\href {\doibase 10.1088/1751-8113/42/50/504008} {\bibfield  {journal}
  {\bibinfo  {journal} {J. Phys.}\ }\textbf {\bibinfo {volume} {A42}},\
  \bibinfo {pages} {504008} (\bibinfo {year} {2009})},\ \Eprint
  {http://arxiv.org/abs/0905.0932} {arXiv:0905.0932 [hep-th]} \BibitemShut
  {NoStop}%
\bibitem [{\citenamefont {Mathur}(2008)}]{Mathur:2008nj}%
  \BibitemOpen
  \bibfield  {author} {\bibinfo {author} {\bibfnamefont {S.~D.}\ \bibnamefont
  {Mathur}},\ }\href@noop {} {\  (\bibinfo {year} {2008})},\ \Eprint
  {http://arxiv.org/abs/0810.4525} {arXiv:0810.4525 [hep-th]} \BibitemShut
  {NoStop}%
\bibitem [{\citenamefont {Papadodimas}\ and\ \citenamefont
  {Raju}(2013)}]{Papadodimas:2012aq}%
  \BibitemOpen
  \bibfield  {author} {\bibinfo {author} {\bibfnamefont {K.}~\bibnamefont
  {Papadodimas}}\ and\ \bibinfo {author} {\bibfnamefont {S.}~\bibnamefont
  {Raju}},\ }\href {\doibase 10.1007/JHEP10(2013)212} {\bibfield  {journal}
  {\bibinfo  {journal} {JHEP}\ }\textbf {\bibinfo {volume} {10}},\ \bibinfo
  {pages} {212} (\bibinfo {year} {2013})},\ \Eprint
  {http://arxiv.org/abs/1211.6767} {arXiv:1211.6767 [hep-th]} \BibitemShut
  {NoStop}%
\bibitem [{\citenamefont {Strominger}\ and\ \citenamefont
  {Vafa}(1996)}]{Strominger:1996sh}%
  \BibitemOpen
  \bibfield  {author} {\bibinfo {author} {\bibfnamefont {A.}~\bibnamefont
  {Strominger}}\ and\ \bibinfo {author} {\bibfnamefont {C.}~\bibnamefont
  {Vafa}},\ }\href {\doibase 10.1016/0370-2693(96)00345-0} {\bibfield
  {journal} {\bibinfo  {journal} {Phys. Lett.}\ }\textbf {\bibinfo {volume}
  {B379}},\ \bibinfo {pages} {99} (\bibinfo {year} {1996})},\ \Eprint
  {http://arxiv.org/abs/hep-th/9601029} {arXiv:hep-th/9601029 [hep-th]}
  \BibitemShut {NoStop}%
\bibitem [{\citenamefont {Sen}(2014)}]{Sen:2014aja}%
  \BibitemOpen
  \bibfield  {author} {\bibinfo {author} {\bibfnamefont {A.}~\bibnamefont
  {Sen}},\ }\href {\doibase 10.1007/s10714-014-1711-5} {\bibfield  {journal}
  {\bibinfo  {journal} {Gen. Rel. Grav.}\ }\textbf {\bibinfo {volume} {46}},\
  \bibinfo {pages} {1711} (\bibinfo {year} {2014})},\ \Eprint
  {http://arxiv.org/abs/1402.0109} {arXiv:1402.0109 [hep-th]} \BibitemShut
  {NoStop}%
\bibitem [{\citenamefont {Chatterjee}\ and\ \citenamefont
  {Lowe}(2018)}]{Chatterjee:2017zeb}%
  \BibitemOpen
  \bibfield  {author} {\bibinfo {author} {\bibfnamefont {A.}~\bibnamefont
  {Chatterjee}}\ and\ \bibinfo {author} {\bibfnamefont {D.~A.}\ \bibnamefont
  {Lowe}},\ }\href {\doibase 10.1088/1361-6382/aab5cc} {\bibfield  {journal}
  {\bibinfo  {journal} {Class. Quant. Grav.}\ }\textbf {\bibinfo {volume}
  {35}},\ \bibinfo {pages} {094001} (\bibinfo {year} {2018})},\ \Eprint
  {http://arxiv.org/abs/1712.03211} {arXiv:1712.03211 [hep-th]} \BibitemShut
  {NoStop}%
\bibitem [{\citenamefont {Parikh}(2004)}]{Parikh:2004rh}%
  \BibitemOpen
  \bibfield  {author} {\bibinfo {author} {\bibfnamefont {M.~K.}\ \bibnamefont
  {Parikh}},\ }in\ \href {\doibase 10.1142/9789812704030_0155} {\emph {\bibinfo
  {booktitle} {{On recent developments in theoretical and experimental general
  relativity, gravitation, and relativistic field theories. Proceedings, 10th
  Marcel Grossmann Meeting, MG10, Rio de Janeiro, Brazil, July 20-26, 2003. Pt.
  A-C}}}}\ (\bibinfo {year} {2004})\ pp.\ \bibinfo {pages} {1585--1590},\
  \Eprint {http://arxiv.org/abs/hep-th/0402166} {arXiv:hep-th/0402166 [hep-th]}
  \BibitemShut {NoStop}%
\bibitem [{\citenamefont {Strominger}(1994)}]{Strominger:1994tn}%
  \BibitemOpen
  \bibfield  {author} {\bibinfo {author} {\bibfnamefont {A.}~\bibnamefont
  {Strominger}},\ }in\ \href@noop {} {\emph {\bibinfo {booktitle} {{NATO
  Advanced Study Institute: Les Houches Summer School, Session 62: Fluctuating
  Geometries in Statistical Mechanics and Field Theory Les Houches, France,
  August 2-September 9, 1994}}}}\ (\bibinfo {year} {1994})\ \Eprint
  {http://arxiv.org/abs/hep-th/9501071} {arXiv:hep-th/9501071 [hep-th]}
  \BibitemShut {NoStop}%
\bibitem [{\citenamefont {Czerniawski}(2006)}]{Czerniawski:2006sc}%
  \BibitemOpen
  \bibfield  {author} {\bibinfo {author} {\bibfnamefont {J.}~\bibnamefont
  {Czerniawski}},\ }in\ \href@noop {} {\emph {\bibinfo {booktitle} {{10th
  International Meeting on Physical Interpretations of Relativity Theory}}}}\
  (\bibinfo {year} {2006})\ \Eprint {http://arxiv.org/abs/gr-qc/0611104}
  {arXiv:gr-qc/0611104} \BibitemShut {NoStop}%
\bibitem [{\citenamefont {Bondi}\ \emph {et~al.}(1962)\citenamefont {Bondi},
  \citenamefont {van~der Burg},\ and\ \citenamefont
  {Metzner}}]{10.2307/2414436}%
  \BibitemOpen
  \bibfield  {author} {\bibinfo {author} {\bibfnamefont {H.}~\bibnamefont
  {Bondi}}, \bibinfo {author} {\bibfnamefont {M.~G.~J.}\ \bibnamefont {van~der
  Burg}}, \ and\ \bibinfo {author} {\bibfnamefont {A.~W.~K.}\ \bibnamefont
  {Metzner}},\ }\href {http://www.jstor.org/stable/2414436} {\bibfield
  {journal} {\bibinfo  {journal} {Proceedings of the Royal Society of London.
  Series A, Mathematical and Physical Sciences}\ }\textbf {\bibinfo {volume}
  {269}},\ \bibinfo {pages} {21} (\bibinfo {year} {1962})}\BibitemShut
  {NoStop}%
\end{thebibliography}%

\end{document}